\documentclass[seceqn]{elsart}

\usepackage{color}
\usepackage{framed}
\usepackage{slashed}
\usepackage{epsfig}

\usepackage{amsfonts}
\usepackage{amsmath}

\usepackage{axodraw}

\newcounter{bla}

 \definecolor{lightgrey}{gray}{0.9}

 \def\btab#1\etab{\begin{tabular}{p{50mm}p{70mm}}#1\end{tabular}}
\def\btabx#1\etabx{\begin{tabular}{p{65mm}p{55mm}}#1\end{tabular}}
\def\btaby#1\etaby{\begin{tabular}{p{20mm}p{100mm}}#1\end{tabular}}
 \def\bcen{\begin{center}}
 \def\ecen{\end{center}}
 
\def\bgfb#1\egfb{\bcen\fcolorbox{black}{lightgrey}{\parbox{130mm}{\btab#1\etab}}\ecen}

\def\bgfbx#1\egfbx{\bcen\fcolorbox{black}{lightgrey}{\parbox{130mm}{\btabx#1\etabx}}\ecen}

\def\bgfbalign#1\egfbalign{\bcen\fcolorbox{black}{lightgrey}{\parbox{130mm}{\btaby#1\etaby}}\ecen}

\newcommand{\eg}[0]{\emph{e.g.{}} }
\newcommand{\ie}[0]{\emph{i.e.{}} }
\newcommand{\etc}[0]{\emph{etc.{}} }
\newcommand{\comment}[1]{}

\begin{document}
\begin{frontmatter}

\begin{flushright}
CP3-08-20\\
MSUHEP-080616
\end{flushright}

\title{FeynRules - Feynman rules made easy}

\author[a]{Neil D. Christensen},
\author[b]{Claude Duhr}

\address[a]{Department of Physics and Astronomy,\\ Michigan State University,\\ East Lansing, MI 48824\\ Email: neil@pa.msu.edu}
\address[b]{Universit\'e catholique de Louvain,\\ Center for particle physics and phenomenology (CP3),\\ Chemin du cyclotron, 2, B-1348 Louvain-La-Neuve, Belgium\\ Email: claude.duhr@uclouvain.be}

\begin{abstract}
In this paper we present FeynRules, a new \emph{Mathematica} package that facilitates the implementation of new particle physics models.  After the user implements the basic model information (\eg particle content, parameters and Lagrangian), FeynRules derives the Feynman rules and stores them in a generic form suitable for translation to any Feynman diagram calculation program.  The model can then be translated to the format specific to a particular Feynman diagram calculator via FeynRules translation interfaces.  Such interfaces have been written for CalcHEP/CompHEP, FeynArts/FormCalc, MadGraph/MadEvent and Sherpa, making it possible to write a new model once and have it work in all of these programs.  In this paper, we describe how to implement a new model, generate the Feynman rules, use a generic translation interface, and write a new translation interface.  We also discuss the details of the FeynRules code.

\begin{flushleft}
\end{flushleft}


\end{abstract}

\end{frontmatter}

\newpage


{\bf PROGRAM SUMMARY}

\begin{small}
\noindent
{\em Manuscript Title: FeynRules - Feynman rules made easy}\\
{\em Authors: Neil D. Christensen, Claude Duhr}\\
{\em Program Title: FeynRules}\\
{\em Journal Reference:}\\
{\em Catalogue identifier:}\\
{\em Licensing provisions:}\\
{\em Programming language: Mathematica}\\
{\em Computer: Platforms on which Mathematica is available.}\\
{\em Operating system: Operating systems on which Mathematica is available.}
{\em Classification: 11.1 General, High Energy Physics and Computing\\
\phantom{Classi}11.2 Phase Space and Event Simulation\\
\phantom{Classi}11.6 Phenomenological and Empirical Models and Theories}\\
{\em Nature of problem: Automatic derivation of Feynman rules from a Lagrangian.\\
Implementation of new models into Monte Carlo event generators and FeynArts.}\\
   \\
{\em Solution method: FeynRules works in two steps: 1) derivation of the Feynman rules directly form the Lagrangian using canonical commutation relations among fields and creation operators. 2) implementation of the new physics model into FeynArts as well as various Monte Carlo programs via interfaces.}\\
   \\
{\em Restrictions: The Lagrangian must fulfill basic QFT requirements, such as Lorentz and gauge invariance. Only fields with spin $0$, $\frac{1}{2}$, $1$ and $2$ are implemented.}\\
   \\
{\em Unusual features: Translation interfaces to FeynArts, CalcHEP/CompHEP, MadGraph and Sherpa exist.}\\
   \\
   \\
{\em Running time: The running time depends on the complexity of the Lagrangian, and varies from seconds (Standard Model) to minutes (more complicated models, like the 3-Site Model).}\\
   \\

\end{small}

\newpage


\begin{center}
\begin{figure}[!h]
\includegraphics[scale=1.2]{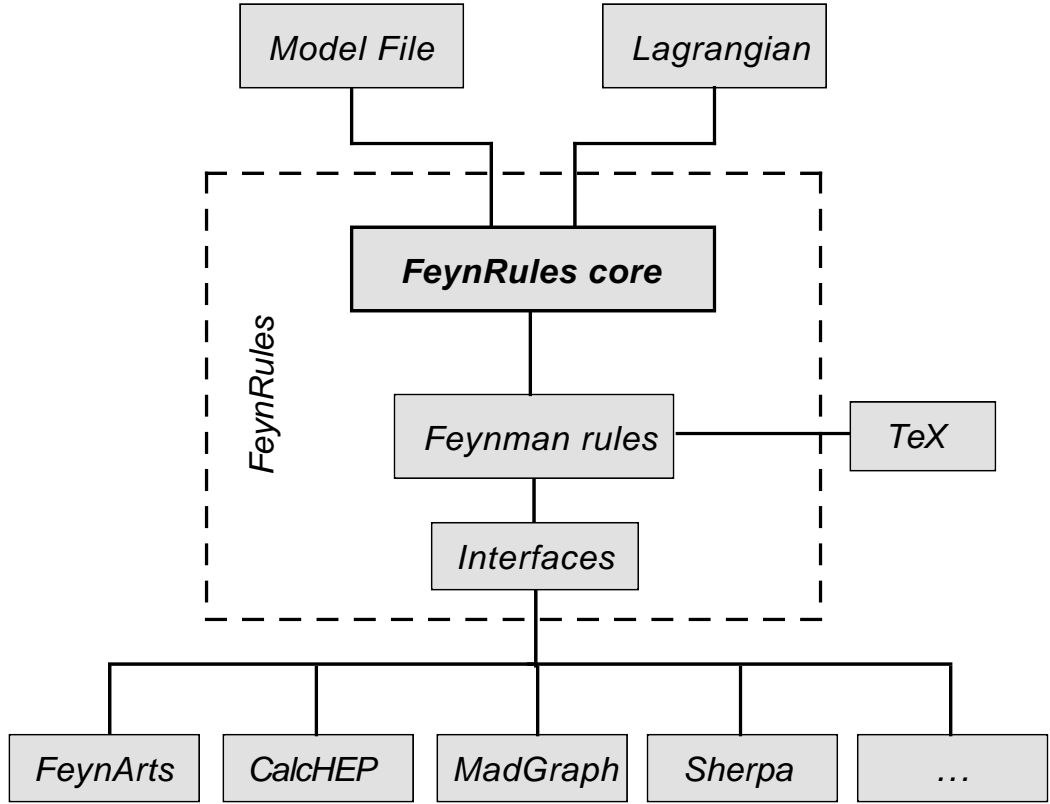}
\caption{The FeynRules flowchart.}
\end{figure}
\end{center}

\tableofcontents

\hspace{1pc}


\section{Introduction}
\label{sec:intro}
The Standard Model (SM) of high energy physics has been amazingly successful at describing and predicting the physics of high energy colliders.  We have found every particle in the SM except the Higgs, a fundamental scalar field which may be responsible for breaking the electroweak symmetry.  We do not know if we will find the Higgs as a fundamental scalar field, but we do know that the SM with the Higgs removed violates unitarity at around a TeV.  On the other hand, even if we discover the Higgs, we expect new physics to emerge at the TeV scale to stabilize it.  Either way, we have strong theoretical reason to believe a rich new spectrum of fields and interactions will emerge at the TeV scale.

The Large Hadron Collider (LHC) will turn on this year with the primary purpose of probing this scale.  To determine which model of new physics is correct requires comparison of the predictions of the models and the results of the experiment.  Such predictions often require the computation of thousands if not tens of thousands of Feynman diagrams.  Such a monumental task is especially suited to computers.  There are several programs on the market that calculate the Feynman diagrams which can be used for this task.  A few of these are CalcHEP/CompHEP\cite{Pukhov:1999gg,Pukhov:2004ca,Boos:2004kh}, FeynArts/FormCalc\cite{Hahn:2000kx,Hahn:1998yk,Hahn:2002vc,Hahn:2004rf,Hahn:2005vh,Hahn:2006qw,Hahn:2006ig,Hahn:2006zy}, Herwig\cite{Corcella:2002jc,Corcella:2000bw,Bahr:2008pv,Bahr:2008tx}, MadGraph/MadEvent\cite{Alwall:2007st,Maltoni:2002qb}, Sherpa\cite{Gleisberg:2003xi}, and Omega/ Whizard \cite{Moretti:2001zz,Kilian:2007gr}.  Each of these programs has its own strengths and it would be ideal if we could simultaneously implement a model in all of these programs.  However, implementing a new model in these Feynman diagram programs can be a tedious and error prone process, often requiring the input of one vertex at a time.  

The first program to take as input a model file with the Lagrangian, derive the Feynman rules and then write the model file for a Feynman diagram calculator was LanHEP\cite{Semenov:1996es,Semenov:1998eb,Semenov:2002jw,Semenov:2008jy}.  It allows the user to write the Lagrangian in a form that is close to the form written on paper, derives all the vertices and then writes a model file appropriate for CalcHEP/CompHEP.  It also allows the user to run several tests on the Lagrangian, thus reducing errors in the implementation.  Nevertheless, this package only produces model files appropriate for CalcHEP/CompHEP, and more recently an output which allows to write FeynArts model files has been added.  A user who wishes to use one of the other Feynman diagram calculators has not been able to use this program directly.

FeynRules is a new package based on \emph{Mathematica}\textregistered\footnote{\emph{Mathematica} is a registered trademark of Wolfram Research, Inc. } which takes a model file with the Lagrangian as input and derives the interaction vertices associated with this Lagrangian. The underlying algorithm, based on canonical quantization formalism, is suitable not only for renormalizable theories, but allows the derivation of the Feynman rules in effective theories involving higher-dimensional operators as well, which makes the package a useful tool for developing models containing the SM as a low-energy effective theory.  Furthermore, this package contains a set of functions allowing the user to test (\eg hermiticity, correct normalization, \etc) and build their model one piece at a time.  In addition to these built-in functions, the user is invited to exploit the full power of the underlying \emph{Mathematica} code to extend these functions and to create his/her own new routines.

The basic input a user provides when implementing his/her model into FeynRules is the so called model file, a text file containing all the properties of the model (particles, parameters, \etc), and the Lagrangian written down using standard \emph{Mathematica} commands, augmented by some new symbols like Dirac matrices, which are necessary when writing down a Lagrangian.
The information contained in the model file, together with the interaction vertices computed inside \emph{Mathematica}, are stored in a generic format which is suitable for any further processing of this information.
In a second step, FeynRules can translate this generic model (with the vertices) into the model format of choice and allows in this way to implement the new model in any tool for which such a ``translation interface'' exists. This approach allows FeynRules to go beyond previous packages in several ways:
\begin{enumerate}
\item FeynRules is not tied to any existing Feynman diagram calculator.
\item The generic model format of FeynRules is suitable to be translated to any other format.
\item The user may choose his favorite Feynman diagram calculator, according to the strength and advantages of the latter.
\item The underlying Mathematica structure allows a `theorist-friendly' environment, which makes the package useful as a sandbox to develop a new model.
\end{enumerate}
Let us comment on some of these points in more detail:

First, as the generic model information is not tied to any specific Feynman diagram calculator or programming language, it is easy to write translation interfaces to other packages without any restrictions on the required format or details of the programming language. In this way, the user can avoid performing different implementations of the same model for different Feynman diagram calculators, which is generally a time-consuming and error prone process. Currently, such interfaces have been written for CalcHEP/CompHEP, FeynArts/FormCalc, MadGraph/MadEvent and Sherpa, and we hope that we will have more interfaces in the future. Furthermore, this approach allows the user to exploit the strength of each Feynman diagram calculator for which a translation interface exists, because once a FeynRules implementation of the model is available, the model can be automatically implemented into any of these codes (so long as the code supports all the vertices in the model).

Second, the user can rely on the power of the underlying \emph{Mathematica} structure when writing the model file and using FeynRules for phenomenological studies. This allows to easily write new user-defined routines without having to rely on knowledge of any programming language exterior to \emph{Mathematica} and provides thus a very general and theorist-friendly environment where extensions of the FeynRules core code can be written. The FeynRules package and core code can always be found together with an up-to-date manual at http://feynrules.phys.ucl.ac.be.

The outline of the paper is as follows: In Section~\ref{sec:modelfile} we discuss the structure of the FeynRules model files, the key ingredient of a FeynRules implementation where all the particles and parameters are declared. In Section~\ref{sec:Lagrangian} we show how to write a Lagrangian in FeynRules using standard \emph{Mathematica} commands, augmented by some new symbols representing special objects like Dirac matrices, and in Section~\ref{sec:simpleexample} we give as a complete example the implementation of QCD with six quark flavors. In Section~\ref{sec:running} we explain how to run the code to derive interaction vertices. Section~\ref{sec:toolbox} is devoted to the ToolBox, a set of additional functions which are not directly related to the computation of the interaction vertices, but which might be useful at different stages during the development and the implementation of a new model. In the Sections~\ref{sec:interfaces} and~\ref{sec:Implementing an Interface} we discuss the translation interfaces to Feynman diagram calculators and how it is possible to write new interfaces using the generic model information contained in FeynRules. Finally, in Section~\ref{sec:codedetails} we briefly discuss some details concerning the underlying algorithms and in Section~\ref{sec:conclusion} we draw our conclusions.

\section{The Model-File}
\label{sec:modelfile}

At the level of Feynman diagram calculations, a new model consists of the following:
\begin{enumerate}
\item a set of quantum fields.
\item a set of parameters.
\item a Lagrangian density.
\end{enumerate}

In order to implement such a new model in a program that calculates Feynman diagrams, this information must be entered in a format appropriate for the respective program.  Each program has its own format making this a tedious and error prone process.  FeynRules solves this problem by allowing the user to write their model in a generic ``FeynRules'' format and then run a FeynRules interface that translates this format into the format appropriate for the Feynman diagram calculation program of choice.  To do this, the user creates a FeynRules model file containing the essential model information as well as various options that simplify the process of writing the Lagrangian and the translation into various Feynman amplitude programs.

Because many fields have similar properties and interactions, it is convenient to group particles into classes.  For this reason, and others, the FeynRules model format was taken to be an extension of the FeynArts format.  This extension includes new classes and new options for each class that are useful for calculating the Feynman rules and/or for translation to a Feynman amplitude program.

In the following subsections, we describe these classes and options and how they are used in the model file.


\subsection*{Model Information}
Information about the model may be included using the optional variables \verb+M$ModelName+\comment{$} and \verb+M$Information+\comment{$}.  The first of these is a string giving the name of the model.  The default model name is the filename of the model file.  

The variable \verb+M$Information+\comment{$} acts as an electronic signature of the authors of the model file. It consists of a replacement list containing the name(s), institution(s), and email(s) of the author(s), the date the model file was created, and a list of references that the author(s) would like  cited when this model file is used. The author list and references will be written to the screen whenever a user loads this model, reminding them to give the authors due credit.  The full contents of \verb+M$Information+\comment{$} can be seen by evaluating the function \verb+ModelInformation[]+ at any time after the model is loaded.  An example of the use of \verb+M$ModelName+\comment{$} and \verb+M$Information+\comment{$} is shown below:
\begin{verbatim}
M$ModelName = "my new model";

M$Information = {Authors -> {"Mr. X", "Ms. Y"},
                 Institutions -> {"UCL"},
                 Emails -> {"X@ucl.be", "Y@ucl.be},
                 Date -> "01. 01. 2010",
                 References -> {"reference 1", "reference 2"}
                };
\end{verbatim}

\begin{table}
\bgfb
\multicolumn{2}{c}{\textbf{Table~\ref{fig:Model Information}: Model Information}}\\
\\
\tt{M\$ModelName} &  A string, specifying the name of the model. The default value is the name of the model file. \\
\tt{M\$Information}  & A replacement list, acting as an electronic signature of the model file.\\
\tt{ModelInformation[ ]} & Prints the contents of {\tt M\$Information} in a \emph{Mathematica} notebook.\\
\multicolumn{2}{l}{\emph{Options for M\$Information}}\\ 
\tt{Authors} & A list of strings, specifying the authors of the model file.\\
\tt{Institutions} & A list of strings, specifying the institutions of the authors.\\
\tt{Emails} & A list of strings, specifying the email addresses of the authors. The order of the email addresses is the same as the order in which the names of the authors have been specified.\\
\tt{Date} & A string specifying the date.\\
\tt{References} & A list of strings, containing the references the authors would liked cited whenever the model implementation is used.
\egfb
\textcolor{white}{\caption{\label{fig:Model Information}}}
\end{table}

 \subsection*{Index Definitions}
Many fields and parameters carry indices specifying their members and how they transform under symmetry groups.  For example, an $SU(N)$ gauge field $G_\mu^a$ carries two indices: 
 \begin{itemize}
\item a Lorentz index $\mu$ ranging from $0$ to $3$.
\item an adjoint gauge index $a$ ranging from $1$ to $N^2-1$.
\end{itemize}
FeynRules treats a field $\psi_{i_1i_2\dots}$ as an object of the form {\tt psi[}$index_1$, $index_2$\ldots{\tt]}, where $index_1$, $index_2$, \ldots denote objects of the form \verb+Index[+\emph{name}, \emph{value}\verb+]+. For FeynRules to run properly, all the indices have to be declared in the model file, together with the range of values they can take. This is done by including the \verb+IndexRange+ command as in the following examples:
\begin{verbatim}
IndexRange[ Index[Colour] ] = Range[3]
IndexRange[ Index[Gluon] ]  = NoUnfold[ Range[8] ]
\end{verbatim}
This declares two indices named \verb+Colour+ and \verb+Gluon+ which range form $1$ to $3$ and $1$ to $8$ respectively. Notice the appearance of the \verb+NoUnfold+ command on the right-hand side of the declaration for the index \verb+Gluon+.  This command is ignored by FeynRules but plays a role in FeynArts.  If included, the FeynArts interface will write this to the FeynArts model file.  For more details on how indices are used inside FeynRules, see Section~\ref{sec:codedetails}.  

As already stated above, an index is represented internally by an object of the form \verb+Index[Colour, ...]+. To make the output more readable, the user can specify how FeynRules should print an index of a given type. This is done via the \verb+IndexStyle+ command.  Some examples are:
\begin{verbatim}
IndexStyle[ Colour, i ]
IndexStyle[ Gauge, a ]
\end{verbatim}
which  tell FeynRules to print indices of type \verb+Colour+ and \verb+Gluon+ as \verb+i+ and  \verb+a+ respectively.

Four-vector indices (\verb+Lorentz+) and Dirac indices (\verb+Spin+) are predefined in FeynRules and do not need to be declared in the model file. Both indices range from $1$ to $4$, and print as $\mu$ and $s$ respectively.

\begin{table}
\bgfb
\multicolumn{2}{c}{\textbf{Table~\ref{fig:Index Information}: Index Information}}\\
\\
{\tt Index[}\emph{name},\emph{value}{\tt]} & Represents an index of type \emph{name} with value \emph{value}.\\
{\tt IndexRange[Index[}\emph{name}{\tt]]} & Declares an index of type \emph{name} along with its range.\\
{\tt Range[}\emph{value}{\tt]} & Declares the range of an index to be from \emph{1} to \emph{value}.\\
{\tt IndexStyle[}\emph{name}, \emph{symbol}{\tt]} & Fixes the {\tt StandardForm} and {\tt TraditionalForm} of an index of type \emph{name} to print as \emph{symbol}.\\
{\tt NoUnfold} & FeynArts command. This is only used by the FeynArts Interface.\\
{\tt Lorentz} & Name for Lorentz indices, ranging from $1$ to $4$, and printing as $\mu$.\\
{\tt Spin} & Name for Dirac indices, ranging from $1$ to $4$, and printing as $s$.
\egfb
\textcolor{white}{\caption{\label{fig:Index Information}}}
\end{table}

\subsection*{Gauge Groups}
\label{sec:gaugegroups}

The structure of a typical Lagrangian is governed by symmetries which fix the form of the interactions. For this reason, it is useful to declare gauge group classes in the FeynRules model file. Here each gauge group class is given a name, and a set of options which specify the details of the gauge group.  All gauge group classes are contained  in the list \verb+M$GaugeGroups+\comment{$}, for example:
\begin{verbatim}
M$GaugeGroups = {
   gaugegroup1 == { options },
   gaugegroup2 == { options },
   ...}
\end{verbatim}\comment{$}
A complete list of options may be found in Table \ref{fig:Gauge Group Options}. We here detail a few of them:

Gauge group classes are divided into abelian and non abelian groups, distinguished by the option \verb+Abelian+ which can take the value \verb+True+ or \verb+False+.

The symbols defined by \verb+SymmetricTensor+ and \verb+ StructureConstant + are internally defined as tensor parameters, which are completely symmetric and antisymmetric in all indices.  In particular, the $SU(2)$ structure constant has been implemented explicitly and is given by the symbol \verb+Eps+, which represents the Levi-Civita totally antisymmetric tensor.

Information about the different representations of each gauge group are specified using the option \verb+Representations+ which is a list of the representations used in the model.  For each representation, a symbol and the index which it acts on can be specified as in
\begin{verbatim}
Representations -> {{T1, Colour1}, {T2, Colour2},...}.
\end{verbatim}
This defines two representations of the gauge group, whose generators are \verb+T1+ and \verb+T2+ acting on the indices \verb+Colour1+ and \verb+Colour2+ respectively. Note that 
\begin{itemize}
\item \verb+T1+ and \verb+T2+ are now automatically defined as tensors with indices {\tt\{Gluon, Colour1, Colour1\}} and {\tt\{Gluon, Colour2, Colour2\}}, where \verb+Gluon+ denotes the index of the adjoint representation, carried by the gauge boson specified by the \verb+GaugeBoson+ option.
\item
The indices \verb+Colour1+ and \verb+Colour2+ as well as \verb+Gluon+ \emph{must} be declared as indices. 
\end{itemize}

As an example, here is the declaration of a $U(1)$ and an $SU(3)$ gauge group:
\begin{verbatim}
M$GaugeGroups ={    ...,
    U1EM == {Abelian -> True,
             GaugeBoson -> A,
             Charge -> Q,
             CouplingConstant -> ee},
                    ...,
    SU3C == {Abelian -> False,
             GaugeBoson -> G,
             StructureConstant -> f,
             CouplingConstant -> gs,
             Representations -> {T, Colour}},
                    ...
    }
\end{verbatim}
where \verb+T+ is the symbol representing the $SU(3)$ matrices in the fundamental representation, and \verb+Colour+ is the name of the gauge index carried by the quarks.  

We note here that when a gauge group is declared, FeynRules automatically constructs the field strength tensor associated with it. This can be used in the Lagrangian, for example, to create the kinetic and self interaction terms for the gauge bosons. For an abelian gauge group, the field strength tensor is given by \verb+FS[A, mu, nu]+, where \verb+A+ is the gauge boson and \verb+mu+ and \verb+nu+ denote the Lorentz indices carried by the field strength tensor.  In the case of a non abelian gauge group, it is given by \verb+FS[G,mu,nu,a]+, where \verb+G+ is the gauge boson, \verb+mu+ and \verb+nu+ denote the Lorentz indices and \verb+a+ the (adjoint) gauge-index carried by the field strength tensor. Although there is a sign convention for the coupling constant, it must be used consistently throughout the Lagrangian. We note that the convention in the field strength tensor defined here is the following: 
\begin{equation}
F_{\mu\nu}^a=\partial_\mu G_\nu^a-\partial_\nu G_\mu^a+g_sf^{abc} G_\mu^bG_\nu^c
\end{equation}
where $g_s$ is the coupling constant, $f$ is the structure constant and $G$ is the gauge boson. This convention can however be changed by setting the variable {\tt FR\$DSign} to -1 (The default value is 1).

\begin{table}
\bgfb
\multicolumn{2}{c}{\textbf{Table~\ref{fig:Gauge Group Options}: Gauge Group Options}}\\
\\
{\tt Abelian} & Mandatory. Specifies whether the gauge group is abelian ({\tt True}) or not ({\tt False}).\\
{\tt GaugeBoson} & Mandatory. Gives the name of the gauge boson associated with the gauge group. The gauge boson name must be included in the particle list. The gauge bosons corresponding to different gauge groups should always be defined in separate particle classes.\\
{\tt CouplingConstant} & Mandatory. Gives the symbol of the coupling constant associated with the gauge group. This symbol must be included in the parameter list.\\
{\tt Charge} & Mandatory for abelian groups. The symbol corresponding to the $U(1)$ charge connected with this gauge group.\\
{\tt StructureConstant} & Mandatory for non abelian groups. The symbol associated with the structure constant of the gauge group.\\
{\tt SymmetricTensor} & The symbol associated with the completely symmetric tensor of a non abelian gauge group.\\
{\tt Representations} & A list, containing all the representations defined for this gauge group. A representation is a list of two elements, {\tt \{T, Colour\}}, {\tt T} being the symbol by which the generators of the representation are denoted, and {\tt Colour} being the name of the index this representation acts on.\\
{\tt Definitions} & A list of replacement rules that should be applied by FeynRules before calculating vertices.\\
\egfb
\textcolor{white}{\caption{\label{fig:Gauge Group Options}}}
\end{table}
\begin{table}
\bgfb
\multicolumn{2}{c}{\textbf{Table~\ref{fig:Field Strength Tensors}: Field Strength Tensors}}\\
\\
{\tt FS[}$A$, $\mu$, $\nu${\tt ]} & The field strength tensor $F_{\mu\nu}$ connected with the $U(1)$ gauge boson $A$.\\
{\tt FS[}$G$, $\mu$, $\nu$, $a${\tt ]} & The field strength tensor $F_{\mu\nu}^a$ connected with the non abelian gauge boson $G$.
\egfb
\textcolor{white}{\caption{\label{fig:Field Strength Tensors}}}
\end{table}

\subsection*{Parameters}
\label{sec:params}
A model also contains many parameters such as coupling constants, mixing angles, masses, gauge charges, \etc.  These are specified in the model file as a list of parameters with options as in the following example:
\begin{verbatim}
M$Parameters = {
     ...,
     param1 == { options },
     param2 == { options },
     ...
     }
\end{verbatim}\comment{$}
where \verb+param1+ and \verb+param2+ are the user-defined names of these parameters and \verb+options+ is a list of optional information about each parameter.  

The first option a user should specify for any parameter is whether it is external (independent) or internal (dependent).  This is done with the {\tt ParameterType} option which can be set to either:
\begin{enumerate}
\item External:  This is an independent parameter and is given by a numerical value (\eg the strong coupling $\alpha_s=0.118$).  This is the default for scalar parameters.
\item Internal:  This is a dependent parameter and is specified in terms of the other parameters (\eg  $g_s = \sqrt{4\pi \alpha_s}$).  This is the default for tensor parameters.
\end{enumerate}
Correctly specifying the relationship between parameters in this way is crucial for obtaining correct results from the Feynman diagram calculators.  For example, it is well known that $M_Z$, $M_W$ and $\cos \theta_W$ are related by $\cos\theta_W=M_W/M_Z$ (at tree level).  Violations of this relation (at tree level) in the SM lead to a loss of unitarity and incorrect results.

FeynRules further distinguishes between two types of parameters:
\begin{enumerate}
\item Scalar parameters which do not carry any indices (\eg coupling constants, masses, \etc).
\item Tensor parameters which carry one or more indices (mixing matrices, gauge matrices, \etc).
\end{enumerate}
A list of the options common to scalar and tensor parameters can be found in Table \ref{fig:Parameter Options} while a list of the options unique to tensor parameters can be found in Table \ref{fig:Tensor Parameter Options}.  

Many Feynman diagram calculators have the strong and electromagnetic interactions built in.  For this reason, it is necessary to use fixed names for the parameters associated with these interactions.  A list of the special parameters is given in Table \ref{fig:Special Parameter Names}.  In the rest of this subsection we comment on the details of scalar and tensor parameters.

\begin{table}
\bgfb
\multicolumn{2}{c}{\textbf{Table~\ref{fig:Parameter Options} : Parameter Options}}\\
{\tt ParameterType} & Specifies whether a parameter is {\tt External} or {\tt Internal}. The default value is {\tt External} for scalar parameters and {\tt Internal} for tensor parameters.\\
{\tt Value} & Gives the value or the formula which defines the parameter. The formula can be in terms of other parameters, whether internal or external. The default value is 1.\\
{\tt Definitions} & A list of replacements rules that should be applied by FeynRules before calculating vertices.\\
{\tt ComplexParameter} & Defines whether the parameter should be treated as real ({\tt False}) or complex ({\tt True}). By default, this option is set to {\tt False} for scalar parameters but to {\tt True} for tensor parameters.\\
{\tt TeX} & This option tells FeynRules how to write the \TeX{} form of this parameter. This is used when writing \TeX{} output.   By default, this is the same as the \emph{Mathematica} symbol.\\
{\tt Description} & A string, describing the parameter.\\
\multicolumn{2}{l}{\emph{Additional information for Feynman diagram calculators}}\\
{\tt ParameterName} & Specifies what to replace the symbol by before writing out the Feynman diagram calculator model files. It should not contain special symbols. By default, this is the same as the \emph{Mathematica} symbol.\\
{\tt BlockName} & Specifies the LH block name of the parameter. By default, the block name is {\tt FRBlock}. This option is only available for external parameters.\\
{\tt OrderBlock} & The LH block number. By default, this number is given by the order in which the parameters appear in the model file, starting from $1$. This option is only available for external parameters.\\
{\tt InteractionOrder} & Specifies the order of the parameter in a particular coupling. This option has no default value.
\egfb
\textcolor{white}{\caption{\label{fig:Parameter Options}}}
\end{table}

\subsubsection*{Scalar parameters}
The value of a scalar parameter is specified by the \verb+Value+ option. For external parameters, the value is a number, whereas for internal parameters, the value is the formula which defines the parameter. The formula can be in terms of other parameters, whether internal or external. However, formulas using other internal parameters should come later in the parameter list. (For example, if the definition of the weak mixing angle is in terms of the W boson mass, then the weak mixing angle definition should come later in the parameter list than the W boson mass definition.) The default value is $1$. Some examples are:
\begin{itemize}
\item \verb+Value -> 127.9+,
\item \verb+Value -> Sqrt[1-sw^2]+,
\item \verb+Value -> Sqrt[ 4 Pi \[Alpha]S ]+.
\end{itemize}

The definition option can be used in place of the value option if desired.  If this is done, the parameter will be replaced according to the definition before the vertices are derived.

The option \verb+ComplexParameter+, which  can take the values \verb+True+ or \verb+False+, determines whether FeynRules treats the parameter as complex. The default value is \verb+False+ (it treats the parameter as real).

Besides these options which are directly used by FeynRules, there is an additional set of options that is not necessary for FeynRules to derive Feynman rules, but is needed by some of the interfaces to Feynman diagram calculators.

Since the parameter can be a general \emph{Mathematica} symbol which may not be appropriate for programming languages other than \emph{Mathematica}, {\tt Parameter- Name} specifies what to replace the symbol by before writing out the Monte Carlo model files. It should not contain special symbols. By default, {\tt Parameter- Name} is the same as the \emph{Mathematica} symbol for the parameter. Some examples are
\begin{itemize}
\item \verb+ParameterName -> aS+,
\item \verb+ParameterName -> lam+
\end{itemize}
for the symbols $\alpha_S$ and $\lambda$ respectively.

Many Monte Carlo programs use a \emph{Les Houches (LH)} format to input the external parameters. The LH block to which a parameter should be added can be chosen via the \verb+BlockName+ option. If no LH block is specified, then the parameter is automatically assigned to the block \verb+FRBlock+. The number by which the external parameter should be labeled inside a given LH block is by default the order in which the parameters appear inside the model file, starting from $1$. This number can however be changed using the \verb+OrderBlock+ option.

Some Monte Carlo programs, such as MadGraph/MadEvent, require that the order of a coupling is known a priori (\eg $g_{s}$ is a coupling of order one in QCD while $\alpha_{s}$ is a coupling of order two). This information can be passed into FeynRules via the \verb+InteractionOrder+ option, which is used in the following way:
\begin{verbatim}
InteractionOrder -> {QCD, 2}
\end{verbatim}
Notice that this option is available for both external and internal parameters, and that it is \emph{not} inferred through the relation among external and internal parameters. Furthermore, it has \emph{no} default value, \ie if left unspecified by the user, the interaction order for this parameter will be left undefined.

We include an example of a complete scalar parameter specification: 
\begin{verbatim}
gs == {
        ParameterType -> Internal,
        Value -> Sqrt[4 Pi \[Alpha]S],
        ParameterName -> G,
        InteractionOrder -> {QCD, 1},
        TeX -> Subscript[g, s],
        Description -> "Strong coupling constant"} 
\end{verbatim}

\subsubsection*{Tensor parameters}
Tensor parameters are declared by giving the option \verb+Indices+ a value as in the following examples:
\begin{itemize}
\item \verb+Indices -> {Index[Generation]}+,
\item \verb+Indices -> {Index[Generation],Index[Generation]}+,
\item \verb+Indices -> {Index[SU2],Index[SU2]}+.
\end{itemize}

Tensors share common options with scalar parameters, but value and/or definitions must be given for each component of the tensor.  For example, the Cabibbo matrix might have the following value option:
\begin{verbatim}
Value -> {CKM[1,2] -> Sin[cabi],
          CKM[1,1] -> Cos[cabi],
          CKM[2,1] -> -Sin[cabi],
          CKM[2,2] -> Cos[cabi]}
\end{verbatim}

Unlike scalar parameters, tensors are by default complex and internal. Tensors also have some new options with respect to scalar parameters.
In many situations, tensors correspond to unitary, hermitian or orthogonal matrices. This can be specified by turning the \verb+Unitary+, \verb+Hermitian+ or \verb+Orthogonal+ options to \verb+True+. The default value for each is \verb+False+.

FeynRules usually assumes the summation convention on indices.  This requires the indices to come in pairs.  However, it is sometimes useful to break this rule.  An example is a diagonal Yukawa matrix.  The summation convention requires the Yukawa interaction term to be written as $H\bar{\psi}_f y_{ff'}\psi_{f'}$.  However, if the Yukawa matrix is diagonal, it is convenient to write this as $y_f H\bar{\psi}_f\psi_f$ which violates the summation convention.  This can be accomplished in FeynRules by setting the option \verb+AllowSummation+ to {\tt True} for the Yukawa coupling $y$.  FeynRules will then allow its index to be summed over along with the two that define the summation convention (in this case the indices on the fermions).  This option is only available for tensors carrying one single index.

A complete tensor specification example is given here:
\label{CKM definition}
\begin{verbatim}
CKM == {
       Indices -> {Index[Generation], Index[Generation]},
       Unitary -> True,
       Definitions -> {CKM[3, 3] -> 1,
                       CKM[i_, 3] :> 0 /; i != 3,
                       CKM[3, i_] :> 0 /; i != 3},
       Value -> {CKM[1,2] -> Sin[cabi],
                 CKM[1,1] -> Cos[cabi],
                 CKM[2,1] -> -Sin[cabi],
                 CKM[2,2] -> Cos[cabi]},
       Description -> "CKM-Matrix"} 
\end{verbatim}

We note the simultaneous use of the Value and the Definitions options.  The definitions are applied before calculating the vertices and thus allow the Feynman rules to be simplified by omitting any vertices proportional to CKM[1,3] and CKM[2,3], which are zero.  The result of this is to speed up both FeynRules and the programs that use the Feynman rules thus generated.  If, on the other hand, all the values had been entered using the Value option, vertices between the first and third generation would have appeared in the vertex list where the coupling would have been zero.  The result would be that the Feynman diagram calculation programs would include these vertices in their calculations and only set them to zero after wards.

\begin{table}
\bgfb
\multicolumn{2}{c}{\textbf{Table~\ref{fig:Tensor Parameter Options}: Tensor parameter options}}\\
{\tt Indices} & Mandatory. A list of indices attached to the tensor.\\
{\tt Unitary} & Defines a tensor as unitary ({\tt True}) or not ({\tt False}). {\tt False} by default.\\
{\tt Hermitian} & Defines a tensor as hermitian ({\tt True}) or not ({\tt False}). {\tt False} by default.\\
{\tt Orthogonal} & Defines a tensor as orthogonal ({\tt True}) or not ({\tt False}). {\tt False} by default.\\
{\tt AllowSummation} & See the description above. {\tt False} by default. This option is only available for tensors with exactly one index.
\egfb
\textcolor{white}{\caption{\label{fig:Tensor Parameter Options}}}
\end{table}

\begin{table}
\bgfb
\multicolumn{2}{c}{\textbf{Table~\ref{fig:Special Parameter Names}: Special Parameter Names}}\\
{\tt gs} & The symbol for the strong coupling constant.\\
{\tt aS} & The {\tt ParameterName} for $g_s^2/4\pi$.\\
{\tt T} & The symbol for the fundamental representation of the strong gauge group.\\
{\tt f} & The symbol for the structure constants of the strong gauge group.\\
{\tt ee} & The symbol for the electromagnetic coupling constant.\\
{\tt aEW} &The {\tt ParameterName} for $e^2/4\pi$.\\
{\tt aEWM1} & The {\tt ParameterName} for $\alpha_{EW}^{-1}$.\\
{\tt Q} & The symbols by which the electric charge is represented.
\egfb
\textcolor{white}{\caption{\label{fig:Special Parameter Names}}}
\end{table}

\subsection*{Particle Classes}
\label{sec:parts}
The declaration of the particle classes in the model file follows similar lines as the parameters and the gauge groups, the main difference being that the classes are labeled according to the spin of the particle, following the original FeynArts syntax:
\begin{itemize}
\item \verb+S+: Scalar fields.
\item \verb+F+: Dirac and Majorana spinor fields.
\item \verb+V+: Vector fields.
\item \verb+T+: Spin-2 fields (only real fields are supported).
\item \verb+U+: Ghost fields (only complex ghosts are supported).
\end{itemize}
Each particle class can contain all the particles with the same quantum numbers (although they may have different masses).  This allows the user to write compact expressions for the Lagrangian, for example, consider the QCD Lagrangian:
\begin{equation}
\mathcal{L}_{QCD}=-\frac{1}{4}G_{\mu\nu}^aG_a^{\mu\nu}+\bar q_f i\slashed{\partial} q_f + g_s \bar q_f\gamma^{\mu}T^a q_f G_\mu^a,
\end{equation}
where $q_f$ denotes the ``quark class'', and avoids writing out explicitly the Lagrangian term for each quark flavor.

The particle declaration is written as:
\begin{verbatim}
M$ClassesDescription = {
   particle1 == { options },
   particle2 == { options },
   ...}
\end{verbatim}\comment{$}
The particle classes have two mandatory options:
\begin{enumerate}
\item Each particle class must be given a name, specified by the \verb+ClassName+ option. This name is the symbol by which the particle class is denoted in the Lagrangian. Note that FeynRules only reads in particle classes  for which \verb+ClassName+ is defined.
\item By setting the option \verb+SelfConjugate+, the user specifies whether the particle has an antiparticle or not.  The possible values for this option are \verb+True+ or \verb+False+.
\end{enumerate}
In addition to these two options, particle classes have several additional properties, which as in the case of the parameter classes can be divided into two classes, those that are used directly by FeynRules and those that are only used by the interfaces to Feynman diagram calculators.  We begin by describing those properties directly related to FeynRules.

Quantum fields are tensors under various symmetry groups and therefore have a collection of indices.  Additionally, a field may carry indices as labels (which may not correspond to symmetry groups) as in the case of generation number.  The Lorentz and Spin indices are automatically declared by FeynRules and do not need to be declared by the user.  Each additional index is specified with the option \verb+Indices+.  Two examples of index declarations are given here:
\begin{verbatim}
Indices -> {Index[Generation]}
Indices -> {Index[Generation], Index[Colour]} 
\end{verbatim}
The order in which the indices are declared is the same as the order in which they are called in the construction of the Lagrangian.

In addition to the tensor indices, a quantum field may carry charges of abelian groups.  These are specified using the \verb+QuantumNumbers+ option. For example
\begin{verbatim}
QuantumNumbers -> {Q -> -1, LeptonNumber -> 1}
QuantumNumbers -> {Q -> 2/3}  
\end{verbatim}

The members of the particle class (which carry the same indices and quantum numbers) are given with the \verb+ClassMembers+ option. If the class contains only one member, this is by default the \verb+ClassName+. Some examples are:
\begin{verbatim}
ClassMembers -> {e, mu, tau}
ClassMembers -> {u, c, t} 
\end{verbatim}
for the charged leptons and charge $+2/3$ quarks respectively.  If the class contains more than one member, the field carries an index which distinguishes the different flavors living in the same class.  This index must be declared using the \verb+Indices+ option as above, but it also must be declared using the \verb+FlavorIndex+ option.  This is necessary so that FeynRules can expand the Lagrangian in this index when generating Feynman rules.  For example:
\begin{verbatim}
FlavorIndex -> Generation
\end{verbatim}

Feynman rules are often desired in the mass eigenbasis, however it is often simpler to write the Lagrangian in terms of the gauge eigenbasis.  FeynRules allows the user to define both sets of fields and relate them using the option \verb+Definitions+.  When FeynRules derives the Feynman rules, it will first apply the definitions, thus transforming the gauge basis into the mass basis.  As an example, consider the relation between the hypercharge gauge boson and the photon and Z bosons:
\begin{verbatim}
Definitions -> {B[mu_] -> -sw Z[mu] + cw A[mu]}
\end{verbatim}

Finally, the masses and decay rates of the different class members can be fixed using the \verb+Mass+ and \verb+Width+ options\footnote{In the following we only discuss Mass. Width works in exactly the same way.}.  \verb+Mass+ is a list of the masses for each class member. If no value is given, the user may specify only the symbol as in:
\begin{verbatim}
Mass -> {MW}
Mass -> {MU,MC,MT}
Mass -> {Mu,MU,MC,MT}
\end{verbatim}
where in the final example, the symbol \verb+Mu+ is given for the entire class, while the symbols \verb+MU+,\verb+MC+ and \verb+MT+ are given to the members.  Notice that the mass symbol \verb+Mu+ for the entire class is by default a tensor parameter, with the {\tt AllowSummation} property set to {\tt True} (See the example in Section~\ref{sec:simpleexample}). If no numerical value is given, a default value of 1 is defined by FeynRules.  The user can specify numerical values for the masses by expanding each symbol into a two-component list as in the following examples:
\begin{verbatim}
Mass -> {MW, Internal} 
Mass -> {MZ, 91.188}
Mass -> {{MU,0}, {MC,0}, {MT, 174.3}}
Mass -> {Mu, {MU, 0}, {MC, 0}, {MT, 174.3}}
\end{verbatim}
In the first example, \verb+MW+ is given the value \verb+Internal+.  This instructs FeynRules that the mass is an internal parameter that is defined by the user in the parameter list.  This is the only case in which a user needs to define a mass in the parameter list.  All other masses given in the definition of the particles should not be defined in the parameter list.

The phase $\phi$ carried by a Majorana fermion $\lambda$ can be specified using the {\tt MajoranaPhase} property for the fermion classes. After it has been declared in the model file, the phase can be called inside a Mathematica program using the command

$\phi$ {\tt = MajoranaPhase[} $\lambda$ {\tt]}

For ghost particles, there is an option \verb+Ghost+ which tells FeynRules the name of the gauge boson this ghost field is connected to. There is a similar option \verb+Goldstone+ for scalar fields.

In addition to the options just described, there is a set of options which are not necessary for the derivation of Feynman rules but which are used by the interfaces to Feynman diagram calculators.  

As in the case of the parameters, the \emph{Mathematica} symbol defined for a particle may not be appropriate for Feynman diagram calculators.   For this reason, the options \verb+ParticleName+ and \verb+AntiParticleName+ allow the user to specify the string (or list of strings) that should be used in the external programs.  An example follows:
\begin{verbatim}
ParticleName -> {"ne","nm","nt"}
\end{verbatim}
where the class members may have been ${\nu_e,\nu_\mu,\nu_\tau}$.

As previously mentioned, it is often convenient to define fields which are not mass eigenstates (for example, gauge eigenstates).  Since these fields are not necessary for the Feynman diagram calculators, there is an option \verb+Unphysical+ which can be set to \verb+True+ and instructs the interface to not write the field to the Feynman diagram calculator model file.

According to the \emph{Les Houches accord}, each particle is represented by a numerical code, called the PDG code of the particle.  This can be specified in the model file via the corresponding option called \verb+PDG+, whose value is the PDG number of the particle, or a list of the PDG numbers of the class members.  An automatically generated PDG code is assigned if this options is omitted. It is however \emph{strongly} recommended to use the existing PDG codes whenever possible.  Two examples should suffice:
\begin{verbatim}
PDG -> 23
PDG -> {12,14,16}
\end{verbatim}

Many Feynman diagram calculators draw the Feynman diagrams they generate.  Some of these allow the user to specify how to draw and label the propagators.  FeynRules allows the user to specify this information with the following options.  The string attached to the propagator is given by \verb+PropagatorLabel+, whose default value is the same as the name of the particle.  If there is more than one member of the class, the option should be set to a list of the strings for the members of the class.  The option \verb+PropagatorArrow+ determines whether an arrow should be placed on the propagator.  The default value is \verb+False+.  The option \verb+PropagatorType+  can take any of the following values:
\begin{itemize}
\item \verb+ScalarDash+: A straight dashed line.
\item \verb+Sine+: A sinusoidal line.
\item \verb+Straight+: A straight solid line.
\item \verb+GhostDash+: A dashed line.
\item \verb+Curly+: A curly (gluonic) line.
\end{itemize}

\begin{table}
\bgfb
\multicolumn{2}{c}{\textbf{Table~\ref{fig:Particle Class Options}: Particle Class Options}}\\
{\tt S}, {\tt F}, {\tt V}, {\tt U}, {\tt T} & Particle classes. \\
{\tt ClassName} & Mandatory. This option gives the symbol by which the class is represented.\\
{\tt SelfConjugate} & Mandatory. Takes the values {\tt True} or {\tt False}.\\
{\tt Indices} & The list of indices carried by the field. Note that Lorentz indices ({\tt Lorentz}, {\tt Spin}) are implicit and not included in this list.\\
{\tt FlavorIndex} & The name of the index making the link between the generic class symbol and the class members.\\
{\tt QuantumNumbers} & A replacement rule list, containing the U(1) quantum numbers carried by the class.\\
{\tt ClassMembers} & A list of all the members of a class. If the class contains only one member, this is by default the {\tt ClassName}.\\
{\tt Mass} & A list of the masses for the class members. A mass can be entered as the symbol that represents the mass in the Lagrangian.  Or, it may be a two-component list, the first element being the symbol for the mass, and the second being the numerical value. A generic symbol with default numerical value $1$ is generated if omitted.\\
{\tt Width} & A list of the decay rates for the class members. Similar to {\tt Mass}.\\
{\tt MajoranaPhase} & The Majorana phase of a Majorana fermion.\\
{\tt Definitions} & A list of replacement rules that should be applied by FeynRules before calculating vertices.
\egfb
\textcolor{white}{\caption{\label{fig:Particle Class Options}}}
\end{table}
\begin{table}
\bgfb
\multicolumn{2}{c}{\textbf{Particle Class Options (continued)}}\\
{\tt ParticleName} & A list of strings, corresponding to the particle names as they should appear in the output files for the Feynman diagram calculation programs. By default, this is the same as defined in {\tt ClassMembers}. This name must satisfy the constraints of whatever Feynman diagram calculation package the user wishes to use it with.\\
{\tt AntiParticleName} & Similar to {\tt ParticleName}.\\
{\tt TeXParticleName} & A list of strings. The default is the same as {\tt ParticleName}.\\
{\tt TeXAntiParticleName} & Similar to {\tt TeXParticleName}.\\
{\tt Unphysical} & If {\tt True}, this declares that the field should not be included in the particle list written for another code by a FeynRules interface. The default is {\tt False}.\\
{\tt PDG} & A list of the PDG codes of the particles. An automatically generated PDG code is assigned if this options is omitted.\\
{\tt PropagatorLabel} & A list of strings propagators should be labeled with when drawing Feynman diagrams. The default value is the same as the {\tt ParticleName}.\\
{\tt PropagatorType} & This specifies how to draw the propagator line for this field. The default value is inferred from the class.\\
{\tt PropagatorArrow} & Whether to put an arrow on the propagator ({\tt True}) or not ({\tt False}).  {\tt False} by default.\\
{\tt FullName} & A string, specifying the full name of the particle, or a list containing the names for each class member. By default {\tt FullName} is the same as {\tt ParticleName}.\\
{\tt MixingPartners} & FeynArts option.\\
{\tt InsertOnly} & FeynArts option.\\
{\tt MatrixTraceFactor} & FeynArts option.
\egfb
\textcolor{white}{\caption{\label{fig:Particle Class Options continued}}}
\end{table}

\section{The Lagrangian}
\label{sec:Lagrangian}
Each new model is specified by a Lagrangian, which contains all the information about the interactions among the particles in the model. Working out the mathematical form of these interactions, \ie the Feynman rules, may however be a tedious task. FeynRules uses the information about the particles and parameters, contained in the model file, and derives the Feynman rules directly from the Lagrangian.

The Lagrangian can either be included in the model file or in a \emph{Mathematica} notebook and is entered using ordinary \emph{Mathematica} commands, augmented by some new commands representing special symbols like Dirac matrices needed to write down the Lagrangian. A summary of these commands can be found in Table \ref{fig:Special Symbols for the Lagrangian}.

Quantum fields are represented inside the \emph{Mathematica} expression for the Lagrangian by an object of the form \verb+psi[a,b,c,...]+, where \verb+psi+ is the name of the field as defined in the model file, and \verb+a+, \verb+b+, \verb+c+,... denote the indices carried by the field, appearing in the order in which they were declared in the model file. The same index notation holds true for tensor parameters. We note that:
\begin{enumerate}
\item If the field carries Lorentz indices (\verb+Lorentz+ or \verb+Spin+), which were implicit in the definition of the particle class, then they always appear in the first positions in \verb+psi[a,b,c,...]+.
\item A class \emph{member} does \emph{not} carry the flavor index which makes the link between the class members and the generic class symbol.
\end{enumerate}

If a quantum field is not self conjugate, then FeynRules automatically creates the symbol for the antiparticle by adding \verb+bar+ to the name of the particle, \eg if \verb+e+ denotes the electron field, then \verb+ebar+ denotes the positron field. There are two additional ways to get the names of the antiparticles, by using the commands \verb+HC[e]+ and \verb+anti[e]+. The command \verb+HC+ is actually much more general, and denotes the hermitian conjugate of an object. In this sense, it can also be used to create the hermitian conjugate of an expression, \eg \verb+HC[L]+ denotes the hermitian conjugate of the Lagrangian \verb+L+. Notice in this context that \verb+HC[e]+does not exactly return the hermitian conjugate of the electron field $e$, but rather the field $\bar e$ defined by $\bar e =e^\dagger\gamma^0$.

For anticommuting fields and parameters, the \emph{Mathematica} \verb+Dot+ command should be used, which prevents \emph{Mathematica} from changing the relative order among these fields and assures that the anticommutation is correctly taken into account in the derivation of the Feynman rules.  However, when vectors and/or matrices are multiplied using the \verb+Dot+ command, the matrix multiplication is done by \emph{Mathematica} and the resulting products are reordered unless the matrix multiplication is wrapped inside the \verb+Inner+ command.  As an example, consider the following product and its desired outcome:
\begin{equation}
(\bar u, \bar d).\left(\begin{array}{c} u\\ d\end{array}\right) = \bar u\,u+\bar d\, d,
\end{equation}
where $u$ and $d$ denote the up and down-type quarks. A naive use of the \emph{Mathematica} \verb+Dot+ command leads to the result
\begin{verbatim}
{ubar, dbar}.{u, d}  =  u ubar + d dbar
\end{verbatim}
where \emph{Mathematica} reorders the fermions incorrectly.  This problem can be overcome by using the \verb+Inner+ command in the following way:
\begin{verbatim}
Inner[Dot, {ubar, dbar}, {u, d}]  =  ubar.u + dbar.d
\end{verbatim}
which instructs \emph{Mathematica} to use the \verb+Dot+ command with each resulting product.  Each matrix multiplication must be wrapped in \verb+Dot+.  For example, if the following product is desired:
\begin{equation}
(\bar u, \bar d).\left(\begin{array}{cc}a_{11}&a_{12}\\a_{21}&a_{22}\end{array}\right).\left(\begin{array}{c} u\\ d\end{array}\right),
\end{equation}
where each element is noncommuting, it must be written as:
\begin{verbatim}
Inner[Dot, Inner[Dot, {ubar, dbar} , 
    {{a11, a12} , {a21, a22}}] , {u, d}]
\end{verbatim}
to obtain the correct result.

When entering the Lagrangian, there are typically several indices to take into account.  If there is no ambiguity, FeynRules will restore many of them.  In order to make this clear, we give several equivalent examples with different levels of index suppression.  We demonstrate these by constructing the QCD Lagrangian with up-type quarks.  The Lagrangian is given by:
\begin{equation}\label{eq:QCD Lagrangian}
\mathcal{L}_{QCD} = -\frac{1}{4} G_{\mu\nu}^aG^{\mu\nu}_a  + i \bar u_f \gamma^\mu \partial_\mu u_f + g_s \bar u_f \gamma^\mu T^a u_f G_\mu^a.
\end{equation}
where $f$ denotes the flavor index of the up-type quark ($f\in\{1,2,3\}$).

The most minimal form of the quark interactions is:
\begin{verbatim}
gs uqbar.Ga[mu].T[a].uq G[mu, a]
\end{verbatim}
\verb+uq+ being the symbol of the class of up-type quarks. Its indices include spin, color and generation. All three of these are implicit. However, these could be made explicit like so:
\begin{verbatim}
gs Ga[mu, s, r] T[a, i, j] uqbar[s, f, i].uq[r, f, j] G[mu, a]
\end{verbatim}
There may be times that this is necessary to get indices correctly contracted. We further note that the generations can be separated and even treated differently if desired. If the generations are given unique names in the particle declaration, then these names can be used. Here is an example that separates the generations but leaves the other indices implicit:
\begin{verbatim}
    gs ubar.Ga[mu].T[a].u G[mu, a]
  + gs cbar.Ga[mu].T[a].c G[mu, a] 
  + gs tbar.Ga[mu].T[a].t G[mu, a] 
\end{verbatim}
or, we can separate the generations and make all the indices explicit like so:
\begin{verbatim}
    gs Ga[mu, s, r] T[a, i, j] ubar[s, i].u[r, j] G[mu, a] 
  + gs Ga[mu, s, r] T[a, i, j] cbar[s, i].c[r, j] G[mu, a] 
  + gs Ga[mu, s, r] T[a, i, j] tbar[s, i].t[r, j] G[mu, a] 
\end{verbatim}
Since the symbols \verb+u+, \verb+c+ and \verb+t+ represent specific generations, they do not carry the generation index $f$.

The gluon kinetic term and self interaction terms are given by the square of the field strength tensor. FeynRules has predefined a symbol for the field strength tensor, \verb+FS[G,mu,nu,a]+ where \verb+G+ is the gauge boson (gluon in this case), \verb+mu+ and \verb+nu+ are Lorentz indices and \verb+a+ is the gauge index. Notice that there is a sign convention between the field strength tensor and the covariant derivative. For the sign convention used in \verb+FS+, see Section~\ref{sec:gaugegroups}. This Lagrangian term can be given in FeynRules as:
\begin{verbatim}
-1/4 FS[G, mu, nu, a] FS[G, mu, nu, a]
\end{verbatim}
When these pieces are combined, they are added together and given a name as in this complete example:
\begin{verbatim}
L = -1/4 FS[G, mu, nu, a] FS[G, mu, nu, a]  
           +  I uqbar.Ga[mu].del[uq, mu]  
           +  gs uq.Ga[mu].T[a].uq G[mu, a]
\end{verbatim}
This line may be contained in the model file or it may be given in a \emph{Mathematica} notebook. Either way, after it is specified, it may be used to create Feynman rules.

\begin{table}
\bgfb
\multicolumn{2}{c}{\textbf{Table~\ref{fig:Special Symbols for the Lagrangian}: Special Symbols for the Lagrangian}}\\
{\tt del[}$\phi$, $\mu${\tt ]} & Partial derivative of $\phi$ with respect to the space-time coordinate $x^\mu$.\\
{\tt ME[}$\mu$, $\nu${\tt ]} & Minkowski metric tensor $\eta_{\mu\nu}$.\\
{\tt IndexDelta[}$i$,$j${\tt ]} & Kronecker delta $\delta_{ij}$.\\
{\tt Eps[}a,\ldots, b{\tt ]} & Totally antisymmetric Levi-Civita tensor with respect to the indices a,...,b.\\
{\tt HC[}\emph{expr} {\tt ]} & Hermitian conjugate of \emph{expr}.\\
{\tt CC[}$\psi${\tt]} & The charge conjugate $\psi^c$ of a Dirac field $\psi$.\\
{\tt FS} & Field strength tensor. See Section~\ref{sec:gaugegroups}.\\
{\tt Ga[}$\mu${\tt ]}, {\tt Ga[}$\mu$, $i$, $j${\tt ]} & Dirac Matrix $\gamma^\mu$, $\gamma^\mu_{ij}$.\\
{\tt ProjP}, {\tt ProjP[}$i$, $j${\tt ]} & Projection operator $\frac{1+\gamma^5}{2}$, $\left(\frac{1+\gamma^5}{2}\right)_{ij}$.\\
{\tt ProjM}, {\tt ProjM[}$i$, $j${\tt ]} & Projection operator $\frac{1-\gamma^5}{2}$, $\left(\frac{1-\gamma^5}{2}\right)_{ij}$.\\
{\tt ProjP[}$\mu${\tt ]}, {\tt ProjP[}$\mu$, $i$, $j${\tt ]} & $\gamma^\mu\frac{1+\gamma^5}{2}$, $\left(\gamma^\mu\frac{1+\gamma^5}{2}\right)_{ij}$.\\
{\tt ProjM[}$\mu${\tt ]}, {\tt ProjM[}$\mu$, $i$, $j${\tt ]} & $\gamma^\mu\frac{1-\gamma^5}{2}$, $\left(\gamma^\mu\frac{1-\gamma^5}{2}\right)_{ij}$.\\
{\tt right[}$\psi${\tt ]} & The right-handed part of the fermion field $\psi$, $\frac{1+\gamma^5}{2}\psi$.\\
{\tt left[}$\psi${\tt ]} & The left-handed part of the fermion field $\psi$, $\frac{1-\gamma^5}{2}\psi$.
\egfb
\textcolor{white}{\caption{\label{fig:Special Symbols for the Lagrangian}}}
\end{table}

\section{A Simple Example}
\label{sec:simpleexample}
In this section, we describe the implementation of QCD in FeynRules.  This implementation is complete in the sense that all the lines of the model file are presented in this section.  This model will not use all the features of FeynRules, but it should clarify how to write a new model.  For more advanced examples, one of the models already implemented should be consulted. 

The Lagrangian we consider is the following:
\begin{equation}
\label{eq:LQCD}
\begin{cal}L\end{cal}_{QCD}=-\frac{1}{4}G_{\mu\nu}^aG_a^{\mu\nu}+\bar{q}_f(i\slashed{D}-m_f)q_f -\bar{\eta}^a\partial_\mu D^\mu\eta^a,
\end{equation}
where $G_{\mu\nu}^a$ denotes the gluon field strength tensor, $\eta^a$ the ghost field associated to the gluon, and $f$ runs over all six quark flavor ($u$, $d$, $s$, $c$, $b$, $t$).

\subsection*{Model Information}
Each model file should begin with the model information, which acts as an electronic signature of the model file. Although this information is optional, it can be useful to keep track of modifications made to the file, as well as the references used to fix all the conventions in the Lagrangian, especially if the user of the model file is different form the author of the file. In our case this information could be entered as follows:
\begin{verbatim}
M$Model_Name = "QCD";

M$Information = {
  Authors      -> {"N. Christensen", "C. Duhr"},
  Institutions -> {"Michigan State University", 
                   "Universite catholique de Louvain (CP3)"},
  Emails       -> {"neil@pa.msu.edu", 
                   "claude.duhr@uclouvain.be"},
  Date         -> "June 6, 2008"
};
\end{verbatim}

\subsection*{Indices}
Apart from the model information, the frontmatter of each model file must contain the declaration of all types of indices used inside the model. In QCD, we have three different fields appearing in the Lagrangian, 
\begin{enumerate}
\item the gluon field $G_\mu^a$, carrying two indices, a Lorentz index $\mu$ ranging from 1 to 4 and an adjoint color index $a$ ranging from 1 to 8.
\item the quark field $q_{s,f,i}$, carrying three indices, a spin index $s$ ranging from 1 to 4, a flavor index ranging from 1 to 6 and a fundamental color index $i$ ranging from 1 to 3.
\item the ghost field $\eta^a$, carrying an adjoint color index $a$.
\end{enumerate}
In Section~\ref{sec:modelfile} we mentioned that the Lorentz and spin indices are hard coded into FeynRules and need therefore not to be declared. This leaves us with three indices to be declared at the beginning of the model file in the following way:
\begin{verbatim}
IndexRange[ Index[Colour] ] = Range[3];
IndexRange[ Index[Gluon]  ] = Range[8];
IndexRange[ Index[Flavor] ] = Range[6];
\end{verbatim}

It is useful to instruct FeynRules how to write the indices, in order to make the output produced by FeynRules more readable.
\begin{verbatim}
IndexStyle[Colour, i];
IndexStyle[Gluon , a];
IndexStyle[Flavor, f];
\end{verbatim}

\subsection*{Gauge Groups}
We now turn to the different kinds of classes we need to define in order to implement our model. As QCD is a gauge theory based on the gauge group $SU(3)_c$, we define this gauge group by adding the corresponding class to \verb+M$GaugeGroups+ like this
\begin{verbatim}
M$GaugeGroups = {

  SU3C == {
    Abelian           -> False,
    GaugeBoson        -> G,
    StructureConstant -> f,
    Representations   -> {T, Colour},
    CouplingConstant  -> gs
  }  
};
\end{verbatim}\comment{$}
Notice that we used the conventions presented in Table~\ref{fig:Special Parameter Names} for the symbols related to the gauge group of the strong interactions. Although this convention is not necessary for FeynRules to derive the correct Feynman rules, it is important when running a translation interface to a Monte Carlo program, which have the strong interactions hard coded.

\subsection*{Parameters}
Next we turn to the definition of the parameter classes. The QCD Lagrangian depends on the following set of free parameters:
\begin{enumerate}
\item the values of the quarks masses $m_f$,
\item the value of the strong coupling constant $g_s$.
\end{enumerate}
As we already explained in Section~\ref{sec:modelfile}, masses are not defined as parameter classes, but are properties of the corresponding particles classes. We will therefore only deal with them in the next subsection where we will define the particle classes entering the model file. We are then left with only one free parameter, the strong coupling constant $g_s$. We will follow the convention used in most Monte Carlo programs to define the numerical value of $g_s$ through the relation $g_s = \sqrt{4\pi \alpha_s}$. Hence, we define two new parameters:
\begin{enumerate}
\item an external parameter $\alpha_s$ with numerical value $0.118$.
\item an internal parameter $g_s$ defined by $g_s = \sqrt{4\pi \alpha_s}$.
\end{enumerate}
This can be achieved by adding the corresponding classes to \verb+M$Parameters+:
\begin{verbatim}
M$Parameters = {
    
  \[Alpha]S == {
    ParameterType    -> External,
    Value            -> 0.118,
    ParameterName    -> aS,
    BlockName        -> SMINPUTS,
    InteractionOrder -> {QCD, 2},
    Description      -> "Strong coupling at the Z pole."
  },
  
  gs == {
    ParameterType    -> Internal,
    Value            -> Sqrt[4 Pi \[Alpha]S],
    ParameterName    -> G,
    InteractionOrder -> {QCD, 1},
    Description      -> "QCD coupling"
  }
    
};
\end{verbatim}\comment{$}
Notice that we have to use the {\tt ParameterName} property to fix the way $\alpha_s$ should be printed when writing out model files for Monte Carlo programs, because the symbol \verb+\[Alpha]S+ cannot be interpreted by any other programming language than \emph{Mathematica}. Furthermore, as $g_s$ is defined in terms of $\alpha_s$, $\alpha_s$ appears before $g_s$ in the parameter list.

\subsection*{Particles}
Finally, we have to define the particle classes, which can be done in the following way:
\begin{verbatim}
M$ClassesDescription = {

  F[1] == {
      ClassName     -> q,
      ClassMembers  -> {d, u, s, c, b, t},
      SelfConjugate -> False,
      Indices       -> {Index[Flavor], Index[Colour]},
      FlavorIndex   -> Flavor,
      Mass          -> {MQ, {MD, 0}, {MU, 0}, {MS, 0}, 
                            {MC, 1.25}, {MB, 4.5}, {MT, 174}},
      Width         -> {WQ, {WD, 0}, {WU, 0}, {WS, 0}, 
                            {WC, 0}, {WB, 0}, {WT, 1.6}},
      PDG           -> {1,2,3,4,5,6}
  },
 
  V[1] == {
      ClassName     -> G,
      SelfConjugate -> True,
      Indices       -> {Index[Gluon]},
      Mass          -> 0,
      Width         -> 0,
      PDG           -> 21
  },
  
  U[1] == {
       ClassName        -> \[Eta],
       SelfConjugate    -> False,
       Indices          -> {Index[Gluon]},
       ParticleName     -> "ghG",
       AntiParticleName -> "ghG~",
       Ghost            -> G,
       Mass             -> 0,
       Width            -> 0,
       QuantumNumbers   -> {GhostNumber -> 1}
  }      

};
\end{verbatim}\comment{$}
This defines three particle classes, corresponding to the quarks, the gluons and the ghosts respectively. Some comments are in order:
\begin{enumerate}
\item As indices of type {\tt Lorentz} and {\tt Spin} are implicit in the definition of the particle classes ({\tt F}, {\tt V}), they do not need to be listed in the {\tt Indices} property.
\item The {\tt FlavorIndex} property is used to define which one of the two indices defined by the {\tt Indices} property for the quark class refers to the class members defined for this class.
\item For the quarks, we defined a generic mass symbol {\tt MQ}, as well as six symbols referring to the masses of each individual class member. Notice that the generic symbol {\tt MQ} is defined as a tensor parameter carrying one single index of type {\tt Flavor}, for which the property {\tt AllowSummation} is turned to {\tt True}.
\item Similar to the definition of \verb+\[Alpha]S+, the name \verb+\[Phi]+ for the ghost field is not appropriate for any other programming language than \emph{Mathematica}. We therefore use the {\tt ParticleName} and {\tt AntiParticleName} properties to define names to be used when writing out model files for Monte Carlo programs.
\item The {\tt Ghost} property for the class {\tt U[1]} defines the field $\phi$ to be the ghost field related to the gluon {\tt G}.
\end{enumerate}

\subsection*{Lagrangian}
Having defined all the classes in the model file, we can now turn to the Lagrangian. The QCD Lagrangian has already been discussed in Section~\ref{sec:Lagrangian}, so we will be brief on this point. The Lagrangian reads:
\begin{verbatim}
LGauge = -1/4 FS[G,mu,nu,a] FS[G,mu,nu,a];

LQuark = I qbar.Ga[mu].del[q, mu] 
         + gs qbar.Ga[mu].T[a].q G[mu,a] 
         - MQ[f]qbar[s,f,c].q[s,f,c];

dBRSTG[mu_,a_] := 1/gs ( del[\[Eta][a], mu] 
                      + gs f[a,a2,a3] G[mu,a2] \[Phi][a3] );

LGhost = - gs \[Eta]bar[a].del[dBRSTG[mu,a],mu];


LQCD = LGauge + LQuark + LGhost;
\end{verbatim}
where we defined the ghost Lagrangian making use of the BRST transformation of the gluon. This is an example of how the user can define new \emph{Mathematica} routines that may be useful to simplify the  writing of the Lagrangian inside \emph{Mathematica}. Furthermore, notice the appearance of the mass term {\tt MQ[f]}, which is the generic mass symbol for the quark class defined for the class {\tt q}. The Lagrangian can be either included in the model file, or just be given in the notebook.

We have now all the ingredients to run FeynRules and derive the interaction vertices associated to this particular Lagrangian, which we will describe in the following section.

\section{Running FeynRules}
\label{sec:running}
After the model file is created and the Lagrangian constructed, it can be loaded into FeynRules and the Feynman rules obtained.
\subsection*{Loading FeynRules}
\label{sec:loadFR}
In order to load the FeynRules package, the user must first specify the directory where FeynRules is stored and then load it in the following way:
\begin{verbatim}
$FeynRulesPath = SetDirectory[ <the address of the package> ];
<< FeynRules`
\end{verbatim}\comment{$}

\subsection*{Loading the Model-File}
\label{sec:loadMF}
After the FeynRules package has been loaded\footnote{We note that the user may want to change the current directory of Mathematica at this point.  Otherwise, all FeynRules output may end up in the main FeynRules directory.}, the model can be loaded using the command \verb+LoadModel+ in the following way:
\begin{verbatim}
LoadModel[ < file.fr >, < file2.fr>, ... ]
\end{verbatim}
The model may be contained in one model file or split among several files.  For FeynRules to run properly, the extension of each model file should be \verb+.fr+. 

\subsection*{Extracting the Feynman Rules}
\label{sec:Feynman Rules}
After the model file and the Lagrangian are loaded, the Feynman rules can be extracted using the command \verb+FeynmanRules+.  For the rest of this section, we will use the QCD Lagrangian defined in Eq.~(\ref{eq:LQCD}).  Here is an example of generating the Feynman rules\footnote{Since the vertices list may be very long, it may be desirable to end this statement with a semicolon.}:
\begin{verbatim}
vertsQCD = FeynmanRules[ LQCD ];
\end{verbatim}
The vertices derived by FeynRules are written out on the screen, and also stored internally in the variable \verb+vertsQCD+. The function \verb+FeynmanRules+ has several options, that are described below.

The user can instruct \emph{Mathematica} to \emph{not} write the Feynman rules to the screen with the option \verb+ScreenOutput+ as in this example:
\begin{verbatim}
vertsQCD = FeynmanRules[ LQCD, ScreenOutput -> False];
\end{verbatim}
In this case, the Feynman rules are generated and stored in \verb+vertsQCD+ but are not displayed on screen.

In the two previous examples, the flavors were not expanded.  For example, the preceding commands will only generate a single quark-gluon vertex (q qbar G).  It is often desirable to expand in flavor indices, obtaining separately the vertices (d dbar G, u ubar G, s sbar G, \emph{etc.}).  To do this, use the \verb+FlavorExpand+ option.  This option can be used to specify individual flavor indices to expand over as in \verb+FlavorExpand->Flavor+ where only the \verb+Flavor+ index is expanded over (and not any other flavor indices if defined).  It may specify a list of flavor indices to expand, as in \verb+FlavorExpand->{Flavor,SU2W}+.  Or, it may be used to expand over all the flavor indices as in \verb+FlavorExpand->True+.
  
The list of Feynman rules can be quite long and it may sometimes be desirable to extract just one or a few vertices.  There are several options available that limit the number of vertices constructed.  We list them here:
\begin{itemize}
\item {\tt MaxParticles -> n} instructs {\tt FeynmanRules} to only derive those vertices whose number of external legs does not exceed {\tt n}. \\The option {\tt MinParticles} works in a similar way.
\item {\tt MaxCanonicalDimension -> n} instructs {\tt FeynmanRules} to only derive those vertices whose canonical dimension does not exceed {\tt n}. \\
The option \verb+MinCanonicalDimension+ works in a similar way.
\item \verb+SelectParticles ->{{...}, {...},...}+ instructs \verb+FeynmanRules+ to only derive the vertices specified in the list.  For example, the command:
\begin{verbatim}
FeynmanRules[ LQCD, SelectParticles ->{{G,G,G}, {G,G,G,G}}]
\end{verbatim}
will only derive the three and four-point gluon vertices.
\item \verb+Contains -> { ... }+ instructs \verb+FeynmanRules+ to only derive the vertices which involve all the particles indicated in the list.
\item \verb+Free -> { ... }+ instructs \verb+FeynmanRules+ to only derive the vertices which do not involve any of the particles indicated in the list.
\end{itemize}

\verb+FeynmanRules+, by default, checks whether the quantum numbers that have been defined in the model file are conserved for each vertex. This check can be turned off by setting the option \verb+ConservedQuantumNumbers+  to \verb+False+. Alternatively, the argument of this option can be a list containing all the quantum numbers FeynRules should check for conservation.

The Feynman rules thus constructed are stored internally as a list of vertices where each vertex is a list of two elements.  The first element enumerates all the particles that enter the vertex while the second one gives the analytical expression for the vertex. We illustrate this for the quark-gluon vertex:

\verb+{{{G, 1}, {qbar, 2}, {q, 3}},+ $i g_s\, \delta_{f_2f_3}\, \gamma^{\mu_1}_{s_2s_3}\, T^{a_1}_{i_2i_3}$ \verb+}+

Each particle is given by a two-component list, the first element gives the name of the particle while the second element defines the label given to the indices referring to this particle in the analytical expression.

As the list of vertices derived by FeynRules can be quite long for complicated models, the \verb+SelectVertices+ routine can be useful.
  The shared options are \verb+MaxParticles+, \verb+MinParticles+, \verb+SelectParticles+, \verb+Contains+ and \verb+Free+.  It does not, however, share the options \verb+MaxCanonicalDimension+ and \verb+MinCanonicalDimension+.  It can be invoked as in:
\begin{verbatim}
vertsGluon = SelectVertices[vertsQCD, 
                    SelectParticles->{{G,G,G},{G,G,G,G}}];
\end{verbatim}

It is sometimes convenient to construct the Feynman rules in sections.  For example, suppose the QCD Lagrangian is split into two pieces as in:
\begin{verbatim}
LQCD = LGauge + LQuarks + LGhosts;
\end{verbatim}
The Feynman rules can be constructed all at once as in the previous examples, or they can be constructed separately as in:
\begin{verbatim}
vertsGluon = FeynmanRules[ LGauge ];
vertsQuark = FeynmanRules[ LQuarks ];
vertsGhosts = FeynmanRules[ LGhosts ];
\end{verbatim}
They can later be merged using the function \verb+MergeVertices+ as in:
\begin{verbatim}
vertsQCD = MergeVertices[ vertsGluon, vertsQuark, vertsGhosts ];
\end{verbatim}
This will merge the results obtained for \verb+vertsGluon+, \verb+vertsQuark+, \verb+vertsGhosts+ into a single list of vertices.  If there are two contributions to the same vertex, they will be combined into one vertex.

\subsection*{Writing \TeX\ Output}
It is possible to create a \TeX{} file containing the details of the model and including the vertices just obtained.  To do this, use the function \verb+WriteTeXOutput+ as in:

{\tt WriteTeXOutput[ vertsGluon, vertsQuark, vertsGhosts,} \emph{options} {\tt ]};

where the argument is a list of the vertices desired in the \TeX{} document.  The output file is by default \verb+M$ModelName+\comment{$} with ``.tex'' added, but can be changed to whatever the user wishes using the option \verb+Output+.
If desired, it is also possible to include the Lagrangian into the \TeX{} file by turning the \verb+PrintLagrangian+ option to \verb+True+.

\subsection*{Manipulating Parameters}
\label{sec:updateparams}
The parameters are also an important part of the model and we include several functions to manipulate them.  The numerical values of any parameter can be obtained with the use of the \verb+NumericalValue+ function used in the following way:
\begin{verbatim}
NumericalValue[ Sin[ cabi ]]
\end{verbatim}
where \verb+cabi+ is the Cabbibo angle.  This will return the numerical value of the sine of the Cabbibo angle.  Any other function of the parameters can be placed in \verb+NumericalValue+.

The user may sometimes want to change the value of a subset of external parameters.  For this, we include the function \verb+UpdateParameters+, which, for example, can be used like:
\begin{verbatim}
UpdateParameters[ gs -> 0.118 , ee -> 0.33 ]
\end{verbatim}
where \verb+gs+ and \verb+ee+ are the strong and electromagnetic coupling constants respectively.  

It may sometimes be useful to write and read the parameters to and from a file.  For this, use \verb+WriteParameters+ and \verb+ReadParameters+ respectively.  By default, they write and read the parameters to and from the file named \verb+M$ModelName+\comment{$} with a ``.pars'' appended, but this can be changed with the options \verb+Output+ and \verb+Input+ as in:
\begin{verbatim}
WriteParameters[Output -> "parameters.pars"]
ReadParameters[Input -> "parameters.pars"]
\end{verbatim}
\verb+WriteParameters+ writes out the external and internal parameters (including masses and widths) to the specified file.  The function \verb+ReadParameters+ only reads in the external parameters (including the external masses and widths).  This gives the user another way to change the values of external parameters.  The user can change the values in the parameter file created by \verb+WriteParameters+, and then read it back in using \verb+ReadParameters+.  Any changes made to the internal parameters are currently ignored.

\begin{table}
\bgfb
\multicolumn{2}{c}{\textbf{Table~\ref{fig:Loading Model Files}: FeynRules Commands}}\\
{\tt LoadModel[}\emph{f1.fr}, \emph{f2.fr}, \ldots{\tt ]} & Loads and initializes the model defined by the model files \emph{f1.fr}, \emph{f2.fr},... .\\
{\tt FeynmanRules[}$\mathcal{L}$, \emph{options}{\tt ]} & Calculates the Feynman rules associated with the Lagrangian $\mathcal{L}$.  The options are given in Table \ref{fig:FR and V Options}.\\
{\tt SelectVertices[}\emph{verts}, \emph{options}{\tt ]} & Selects a subset of the vertices contained in \emph{verts}.  The options are given in Table \ref{fig:FR and V Options}.\\
{\tt MergeVertices[}\emph{v1},\emph{v2},\ldots{\tt ]} & Merges the vertex lists \emph{v1}, \emph{v2},\ldots into one list.\\
{\tt WriteTeXOutput[}\emph{v1},\emph{v2},\ldots{\tt ]} & Writes a \TeX{} file containing the model information, particles, parameters and the vertices in the vertex lists \emph{v1}, \emph{v2},\ldots. The name of the \TeX{} file is given by the {\tt Output} option, by default it is the model name with ``.tex'' appended.\\
{\tt NumericalValue[} \emph{f[pars]} {\tt ]} & Gives the numerical value of \emph{f[pars]} where \emph{f} is some function and \emph{pars} is some set of parameters in the model.\\
{\tt UpdateParameters[}{$p1\rightarrow v1$}, $p2\rightarrow v2$,\ldots{\tt ]} & Changes the values of the parameters \emph{p1,p2,\ldots} to \emph{v1,v2,\ldots}.\\
{\tt WriteParameters[}\emph{options}{\tt ]} & Writes the internal and external parameters (including masses and widths) to a text file.  The text file is {\tt M\$ModelName} with ``.pars'' appended unless otherwise specified with the {\tt Output} option.\\
{\tt ReadParameters[}\emph{options}{\tt ]} & Reads the external parameters from a text file.  The text file is named as in {\tt WriteParameters} and must have the same format as created by {\tt WriteParameters}.
\egfb
\textcolor{white}{\caption{\label{fig:Loading Model Files}}}
\end{table}

\begin{table}
\bgfb
\multicolumn{2}{c}{\textbf{Table~\ref{fig:FR and V Options}: FeynmanRules and SelectVertices Options}}\\
{\tt ScreenOutput} & If turned to {\tt False}, the Feynman rules derived will not appear on the screen. The default is {\tt True}.  Just a {\tt FeynmanRules} option.\\
{\tt FlavorExpand} & Expands over the flavor indices specified by the list at the right-hand side of the argument. If turned to {\tt True}, all flavor indices are expanded.  Just a {\tt FeynmanRules} option.\\
{\tt ConservedQuantumNumbers} & Checks whether the quantum numbers specified by the list at the right-hand side of the argument are conserved. If {\tt True} ({\tt False}), all (no) quantum numbers are checked. The default is {\tt True}.  Just a {\tt FeynmanRules} option.\\
{\tt MinParticles} & Only vertices involving at least the specified number of external fields are derived.\\
{\tt MaxParticles} & Only vertices involving at most the specified number of external fields are derived.\\
{\tt MinCanonicalDimension} & Only vertices of at least the specified canonical dimension are derived.  Just a {\tt FeynmanRules} option.\\
{\tt MaxCanonicalDimension} & Only vertices of at most the specified canonical dimension are derived.  Just a {\tt FeynmanRules} option.\\
{\tt SelectParticles} & Calculates only the vertices specified in the list at the right-hand side of the argument.\\
{\tt Contains} & Only the vertices which contain at least the particles contained in the list at the right-hand side of the argument are derived.\\
{\tt Free} & Only the vertices which do not contain any of the particles contained in the list at the right-hand side of the argument are derived.
\egfb
\textcolor{white}{\caption{\label{fig:FR and V Options}}}
\end{table}

\section{The ToolBox}
\label{sec:toolbox}
The FeynRules ToolBox contains a collection of functions that are not directly related to or needed for the derivation of the Feynman rules, but that turn out to be very useful at different stages of the model building and the implementation of the model into FeynRules. This set of functions ranges from useful routines to manipulate the vertex lists produced by \verb+FeynmanRules+ to checking properties of the Lagrangian. Furthermore, the ToolBox is meant to provide the user with an environment where he/she can develop and store his/her own functions written during the implementation of a model. In the rest of this section we are going to describe in more detail the contents of the current version of the ToolBox.

\subsection*{Boolean functions}
In general, writing new \emph{Mathematica} functions for FeynRules requires  identifying which symbols correspond to particles, parameters, \etc. For this reason, the ToolBox contains a set of boolean functions that make this identification. These boolean functions can be used in pattern matching involved in the writing of new routines by the user. A summary of the boolean function available is given in Tables \ref{fig:Boolean Functions for Particles} and \ref{fig:Boolean Functions for Parameters} .

\begin{table}
\bgfb
\multicolumn{2}{c}{\textbf{Table~\ref{fig:Boolean Functions for Particles}: Boolean Functions for Particles}}\\
{\tt FieldQ} & Returns {\tt True} for fields.\\
{\tt FermionQ} & Returns {\tt True} for fermions.\\
{\tt BosonQ} & Returns {\tt True} for bosons.\\
{\tt SelfConjugateQ} & Returns {\tt True} for self conjugate fields.\\
{\tt ScalarFieldQ} & Returns {\tt True} for scalar fields.\\
{\tt DiracFieldQ} & Returns {\tt True} for Dirac fermions.\\
{\tt MajoranaFieldQ} & Returns {\tt True} for Majorana fermions.\\
{\tt VectorFieldQ} & Returns {\tt True} for vector fields.\\
{\tt Spin2FieldQ} & Returns {\tt True} for spin 2 fields.\\
{\tt GhostFieldQ} & Returns {\tt True} for ghost fields.\\
\egfb
\textcolor{white}{\caption{\label{fig:Boolean Functions for Particles}}}
\end{table}
\begin{table}
\bgfb
\multicolumn{2}{c}{\textbf{Table~\ref{fig:Boolean Functions for Parameters}: Boolean Functions for Parameters}}\\
{\tt numQ} & Returns {\tt True} for all parameters in the model, as well as for all numerical values.
               Note that the components of a tensor are considered as parameters, e.g. {\tt numQ[T[a, i, j]] = True}, and   
               {\tt numQ[Ga[mu, r, s]] = True}.\\
{\tt CnumQ} & Returns {\tt True} if a parameter is a complex parameter.\\
{\tt TensQ} & Returns {\tt True} for a tensor.\\
\tt {CompTensQ} & Returns {\tt True} for a complex tensor parameter.\\
{\tt UnitaryQ} & Returns {\tt True} for unitary tensors.\\
{\tt HermitianQ} & Returns {\tt True} for hermitian tensors.\\
{\tt OrthogonalQ} & Returns {\tt True} for orthogonal tensors.\\
\multicolumn{2}{l}{\emph{Additional useful functions}}\\
{\tt MR\$QuantumNumbers} & A list containing all quantum numbers defined in the model file.\\
{\tt \$IndList[}$\psi${\tt ]} & The list of indices declared for the field or the tensor $\psi$.
\egfb
\textcolor{white}{\caption{\label{fig:Boolean Functions for Parameters}}}
\end{table}

\subsection*{Manipulating a Lagrangian}
The FeynRules ToolBox provides the user with a set of functions that are useful at different stages of the model building and the implementation into \emph{Mathematica}.

First, the {\tt ExpandIndices} function allows the user to restore all the suppressed indices in fermion chains, as well as to identify which kind of index a given symbol represents (See Section~\ref{sec:codedetails} for more detail). {\tt ExpandIndices} has the same options as {\tt FeynmanRules}. In particular, the command

{\tt ExpandIndices[} $\mathcal{L}$ {\tt, FlavorExpand -> True]}

will return the Lagrangian $\mathcal{L}$ expanded over all flavor indices.

There is an additional set of functions which allow the user to filter out given sectors of the Lagrangian, \eg

{\tt GetKineticTerms[} $\mathcal{L}$, \emph{options} {\tt]}

where \emph{options} denotes any of the options of {\tt FeynmanRules}, returns all kinetic terms in $\mathcal{L}$ (defined as quadratic terms with derivatives). The functions {\tt GetMassTerms}, {\tt GetQuadraticTerms} and {\tt GetInteractionTerms} work in a similar way. Finally, it is also possible to filter out only those terms of the Lagrangian which correspond to a given interaction (\eg only the three and four-point gluon vertices). This can be achieved using the {\tt SelectFieldContent} function as follows

{\tt SelectFieldContent[} $\mathcal{L}$ {\tt, \{\{G, G, G\}, \{G, G, G, G\}\}]}

Furthermore, FeynRules has the possibility to calculate the values of the masses that appear in the Lagrangian\footnote{Notice that as FeynRules does \emph{not} currently diagonalize mass matrices, the Lagrangian must be in mass diagonal form in order to use this feature.}, both in a numeric and symbolic way. This can be accomplished using the {\tt GetMassSpectrum} function, which has again the same options as {\tt FeynmanRules}.

\begin{table}
\bgfbx
\multicolumn{2}{c}{\textbf{Table~\ref{fig:Manipulating a Lagrangian}: Manipulating a Lagrangian}}\\
\multicolumn{2}{l}{\emph{All the functions introduced in this section have the same options as}}\\
\multicolumn{2}{l}{\tt{FeynmanRules}.}\\
{\tt ExpandIndices[}$\mathcal{L}$, \emph{options}\tt{]} & Restores all the suppressed indices in the Lagrangian $\mathcal{L}$.\\
{\tt GetKineticTerms[}$\mathcal{L}$, \emph{options}{\tt]} & Returns the kinetic terms in the Lagrangian $\mathcal{L}$.\\
{\tt GetMassTerms[}$\mathcal{L}$, \emph{options}{\tt]} & Returns the mass terms in the Lagrangian $\mathcal{L}$.\\
{\tt GetQuadraticTerms[}$\mathcal{L}$, \emph{options}{\tt]} & Returns the quadratic terms in the Lagrangian $\mathcal{L}$.\\
{\tt GetInteractionTerms[}$\mathcal{L}$, \emph{options}{\tt]} & Returns the interaction terms in the Lagrangian $\mathcal{L}$.\\
{\tt SelectFieldContent[}$\mathcal{L}$, \emph{list}\tt{]} & Returns the part of the Lagrangian $\mathcal{L}$ corresponding to the field content specified in \emph{list}.
\egfbx
\textcolor{white}{\caption{\label{fig:Manipulating a Lagrangian}}}
\end{table}

\subsection*{Checking a Lagrangian}
In general, a QFT Lagrangian has to fulfill a set of basic requirements, such as hermiticity, gauge invariance, \emph{etc}. The FeynRules ToolBox contains a set of functions which perform several such checks.

Hermiticity of the Lagrangian can be checked using the command

{\tt CheckHermiticity[} $\mathcal{L}$, \emph{options} {\tt ]}

where \emph{options} denotes any of the options of {\tt FeynmanRules}. Hermiticity is then checked by calculating the Feynman rules of $\mathcal{L} -\mathcal{L}^\dagger$, which of course should all vanish if $\mathcal{L}$ is hermitian.

For FeynRules to work properly, all the quadratic pieces should be diagonal.  We have three commands to check this: \verb+CheckDiagonalQuadraticTerms+, \verb+CheckDiagonalKineticTerms+ and \verb+CheckDiagonalMassTerms+.  An example of how to use these functions is given here:

{\tt CheckDiagonalQuadraticTerms[} $\mathcal{L}$, \emph{options} {\tt]}

where \emph{options} denotes again any of the options of {\tt FeynmanRules}. Furthermore, it is possible to check whether the kinetic terms are correctly normalized\footnote{All kinetic terms must be diagonal to use this function.}, using 

{\tt CheckKineticTermNormalisation[} $\mathcal{L}$, \emph{options} {\tt]}

Finally, FeynRules can check whether the mass spectrum given in the model file corresponds to the masses obtained from the Lagrangian\footnote{All mass terms must be diagonal to use this function.}:

{\tt CheckMassSpectrum[} $\mathcal{L}$, \emph{options} {\tt]}

The normalization that FeynRules assumes for the kinetic terms and the mass terms is the following:
\begin{enumerate}
\item Scalars:
\begin{itemize}
\item[-] Real: 
\begin{equation}
\frac{1}{2}\partial_\mu \phi\partial^\mu\phi -\frac{1}{2}m^2\phi^2,\nonumber
\end{equation}
\item[-] Complex (including ghost fields): 
\begin{equation}
\partial_\mu \phi^\dagger\partial^\mu\phi -m^2\phi^\dagger\phi,\nonumber
\end{equation}
\end{itemize}
\item Spin-1/2 fermions:
\begin{itemize}
\item[-] Majorana: 
\begin{equation}
\frac{1}{2}\bar\lambda i\slashed{\partial}\lambda -\frac{1}{2}m\bar\lambda\lambda,\nonumber
\end{equation}
\item[-] Dirac: 
\begin{equation}
\bar\psi i\slashed{\partial}\psi -m\bar\psi\psi,\nonumber
\end{equation}
\end{itemize}
\item Vectors:
\begin{itemize}
\item[-] Real: 
\begin{equation}
-\frac{1}{4}F_{\mu\nu}F^{\mu\nu}-\frac{1}{2}m^2A_{\mu}A^{\mu},\nonumber
\end{equation}
\item[-] Complex: 
\begin{equation}
-\frac{1}{2}F_{\mu\nu}^\dagger F^{\mu\nu}-m^2A_{\mu}^\dagger A^{\mu},\nonumber
\end{equation}
\end{itemize}
\end{enumerate}

\begin{table}
\bgfbalign
\multicolumn{2}{c}{\textbf{Table~\ref{fig:Checks Lagrangian}: Checking a Lagrangian}}\\
\multicolumn{2}{l}{\emph{All the functions introduced in this section have the same options as}}\\
\multicolumn{2}{l}{\tt{FeynmanRules}.}\\
\multicolumn{2}{l}{{\tt CheckHermiticity[} $\mathcal{L}$, \emph{options} {\tt]}}\\
 & Checks if the Lagrangian $\mathcal{L}$ is hermitian.\\
\multicolumn{2}{l}{{\tt CheckDiagonalKineticTerms[} $\mathcal{L}$, \emph{options} {\tt]}}\\
 & Checks if all the kinetic terms in the Lagrangian $\mathcal{L}$ are diagonal.\\
\multicolumn{2}{l}{{\tt CheckDiagonalMassTerms[} $\mathcal{L}$, \emph{options} {\tt]}}\\
 & Checks if all the mass terms in the Lagrangian $\mathcal{L}$ are diagonal.\\
\multicolumn{2}{l}{{\tt CheckDiagonalQuadraticTerms[} $\mathcal{L}$, \emph{options} {\tt]}}\\
 & Checks if all the quadratic terms in the Lagrangian $\mathcal{L}$ are diagonal.\\
\multicolumn{2}{l}{{\tt CheckKineticTermNormalisation[} $\mathcal{L}$, \emph{options} {\tt]}}\\
 & Checks if all the kinetic terms in the Lagrangian $\mathcal{L}$ are correctly normalized.\\
\multicolumn{2}{l}{{\tt CheckMassSpectrum[} $\mathcal{L}$, \emph{options} {\tt]}}\\
 & Checks if all the mass terms in the Lagrangian $\mathcal{L}$ are correctly normalized and if their value corresponds to the value given in the model file.\\
\egfbalign
\textcolor{white}{\caption{\label{fig:Checks Lagrangian}}}
\end{table}

\subsection*{Manipulating vertex lists}

In Section~\ref{sec:running} we already explained how to use the functions {\tt SelectVertices} and {\tt MergeVertices} to extract individual vertices from the list of results obtained by {\tt FeynmanRules} and to merge different vertex lists into a single one. In this section we introduce two more functions which can be useful for manipulating the vertex lists.

It can sometimes be useful to apply momentum conservation in order to simplify the expression obtained for a vertex, or to rewrite it in an equivalent way. The FeynRules ToolBox provides for this reason a function which allows to replace the momentum of particle number $n$, by minus the sum of all other particles. This is done as follows

{\tt MomentumReplace[} \emph{vertex}, $n$ {\tt ]}

where \emph{vertex} represents a single vertex, \ie an element of a vertex list. 

It is often desirable to simplify an entire set of vertices using momentum conservation.  We have created the function {\tt ApplyMomentumConservation} for this purpose.  For each vertex, it uses {\tt MomentumReplace}, cycling through each momentum and comparing the size of each expression.  It then keeps the shortest expression for the vertex.  It is used as in:

{\tt ApplyMomentumConservation[} \emph{vertexlist} {\tt ]}

\begin{table}
\bgfbx
\multicolumn{2}{c}{\textbf{Table~\ref{fig: Handling vertex lists}: Handling vertex lists}}\\
{\tt SelectVertices[} \emph{vertexlist}, \emph{list} {\tt]} & Returns the vertices in \emph{vertexlist} which match the particle content specified in \emph{list}.\\
{\tt MergeVertices[}\emph{v1},\emph{v2},\ldots {\tt]} & Returns the vertex list obtained by merging the vertex lists \emph{v1}, \emph{v2}, \ldots\\
{\tt MomentumReplace[} \emph{vertex}, $n$ {\tt ]} & Replaces the momentum of particle number $n$ in \emph{vertex} by minus the sum of all other momenta.\\
{\tt ApplyMomentumConservation[}\\\quad \emph{vertexlist}{\tt]} & Applies {\tt MomentumReplace} in all possible ways and returns the results that lead to the simplest vertex.
\egfbx
\textcolor{white}{\caption{\label{fig: Handling vertex lists}}}
\end{table}

\section{Interfaces}
\label{sec:interfaces}
Once Feynman rules have been obtained, the user is typically interested in calculating Feynman diagrams.  There are many programs that do this automatically and each has its own model file format.  FeynRules was constructed to allow easy translation to the various formats available via ``interfaces''.  Translation interfaces have been written for the following Feynman diagram calculators:
\begin{itemize}
\item[-] CalcHEP/CompHEP,
\item[-] FeynArts/FormCalc,
\item[-] MadGraph/MadEvent,
\item[-] Sherpa.
\end{itemize}
Of course, there are many other Feynman diagram calculators and we invite experts in these other Feynman diagram calculators to write translation interfaces for their favorite calculators (see Section \ref{sec:Implementing an Interface}).  

These interfaces are invoked with commands like \verb+Write__Output[l1,l2,...,+ \verb+options]+ where \verb+__+ is replaced with a particular type of interface name.  For example, we have \verb+WriteCHOutput+, \verb+WriteFeynArtsOutput+, \verb+WriteMGOutput+ and \verb+WriteSHOutput+.  \verb+l1,l2,..+ are the Lagrangian terms whose vertices are desired in the output and \verb+options+ are the options specific to each interface.  The details of the interfaces will appear in a later paper and on the FeynRules website\cite{FRwebpage}.  For now, we discuss some general constraints that apply to all the interfaces.

The main limitations of these interfaces are that they can only implement vertices, particles, parameters and names that are allowed by the Feynman diagram calculators that they are being translated to.  Thus, if an interaction is not allowed in the calculator, it will also not be allowed by the interface.  Generally, the interface will warn the user when it comes across a term that is not supported and allow the user to fix it.  In some cases, the interface may fix it on the fly and warn the user what occurred.  A user who wishes to use multiple Feynman diagram calculators will want to satisfy the constraints of each.  We encourage satisfying the constraints of as many Feynman diagram calculators as possible to make your model as widely useful as possible.  In any case, fixing a problematic parameter or particle name, or some other minor constraint violation is not usually difficult.

Furthermore, the implementer of a new model should realize that each diagram calculator works in a certain gauge.  In particular, there are several that work in Feynman gauge and several others that work in unitary gauge.  Since these gauges involve different particle sets and Lagrangian terms, we have found that it is useful to define a switch that can be used in {\tt if}...{\tt else} statements in the model file.  For example, the user might define the variable \verb+FeynmanGauge+ and set it to {\tt True} or {\tt False} depending on which calculation program he/she desires to use.  Then, for example, the Lagrangian terms for the Goldstone bosons could be turned on when \verb+FeynmanGauge+ is turned to {\tt True} and turned off when it is set to {\tt False}.

\subsection*{Special Names}
There are several parameters and particles that have special significance in Feynman diagram calculators.  Examples of these are the strong and electromagnetic couplings, the names of the fundamental representation and the structure constants of the strong gauge group and so on.  For this reason, we have chosen to fix the names of these objects at the FeynRules level.  Adherence to these standards will increase the chances of successful translation.

The strong gauge group has special significance in many Feynman diagram calculators.  A user who implements the strong gauge group should adhere to the following rules.  The indices for the fundamental and adjoint representations of this gauge group should be called \verb+Colour+ and \verb+Gluon+ respectively.  Furthermore, the names of the QCD gauge boson, coupling constant, structure constant, totally symmetric term and fundamental representation should be given by \verb+G+, \verb+gs+, \verb+f+, \verb+dSU3+ and \verb+T+ as in the following example:
\begin{verbatim}
   SU3C == {
       Abelian -> False,
       GaugeBoson -> G,
       StructureConstant -> f,
       SymmetricTensor -> dSU3,
       Representations -> {T, Colour},
       CouplingConstant -> gs
       }
\end{verbatim}
In addition, the strong coupling constant and its square over $4\pi$ should be declared in the parameter section in the following form:
\begin{verbatim}
   \[Alpha]S == {
       ParameterType -> External,
        Value -> 0.118,
        ParameterName -> aS,
        BlockName -> SMINPUTS,
        InteractionOrder -> {QCD, 2},
        Description -> "Strong coupling at the Z pole."
        },
   gs == {
        ParameterType -> Internal,
        Value -> Sqrt[4 Pi \[Alpha]S],
        ParameterName -> G,
        InteractionOrder -> {QCD, 1}
       }
\end{verbatim}
Note that $\alpha_S$ is given as the external parameter and $g_S$ as the internal parameter.  The description of $\alpha_S$ may be edited, but it should be remembered that, for the Monte Carlo programs that run the strong coupling constant, the value of $\alpha_S$ should be set at the Z pole.  For calculation programs that do not run the strong coupling, on the other hand, it should be set according to the scale of the interaction.  A description may also be added to the parameter $g_S$.

The electromagnetic interaction also has special significance in many Feynman diagram calculators and we outline the following standard definitions.  The electric coupling constant should be called \verb+ee+, the electric charge should be called \verb+Q+.  The declaration of the electric charge should follow the following conventions for naming:
\begin{verbatim}
   \[Alpha]EWM1 == {
        ParameterType -> External,
        Value -> 127.9,
        ParameterName -> aEWM1,
        BlockName -> SMINPUTS,
        InteractionOrder -> {QED, -2},
        Description -> "alpha_EM inverse at the Z pole."
        },
   \[Alpha]EW == {
        ParameterType -> Internal,
	        Value -> 1/\[Alpha]EWM1,
	        InteractionOrder -> {QED, 2},
        ParameterName -> aEW,
	        },
   ee == {
        ParameterType -> Internal,
        Value -> Sqrt[4 Pi \[Alpha]EW ],
        InteractionOrder -> {QED, 1}
        }
\end{verbatim}
As for the strong coupling, the description of $\alpha_{EW}^{-1}$ may be edited\footnote{The reason for choosing $\alpha_{EW}^{-1}$ as the external input parameter, and not $\alpha_{EW}$ itself is only to be compliant with the \emph{Les Houches Accord}.}, but it should be remembered that for calculation programs that run the electric coupling, it should be set at the Z pole.  For programs which do not run it, the electric coupling should be set at the interaction scale.  Again, a description may be added to the definition of \verb+\[Alpha]EW+ and \verb+ee+.

The Fermi constant and the Z pole mass are very precisely known and are often used in calculators to define coupling constants and the scale where couplings are run from.  They should be included in the {\tt SMINPUTS} block of the Les Houches accord and should be defined by at least the following:
\begin{verbatim}
   Gf == {
        ParameterType -> External,
        Value -> 1.16639 * 10^(-5),
        BlockName -> SMINPUTS,
        InteractionOrder -> {QED, 2},
        Description -> "Fermi constant"
        },
   ZM == {
        ParameterType -> External,
        Value -> 91.188,
        BlockName -> SMINPUTS,
        Description -> "Z pole mass"
        }
\end{verbatim}

Moreover, the weak coupling constant name \verb+gw+ and the hypercharge symbol \verb+Y+ are used by some calculators and the user is encouraged to use these names where appropriate.  The masses and widths of particles should be assigned whenever possible.  If left out, FeynRules will assign the value 1 to each.  Finally, particles are also identified by a PDG number.  The user is strongly encouraged to use existing PDG codes in their model wherever possible.  If not included, a PDG code will be automatically assigned by FeynRules beginning at 6000001.

The specific details of each interface will be published soon.  They can also be found on the FeynRules website\cite{FRwebpage}.


\section{Implementing an Interface}
\label{sec:Implementing an Interface}
The philosophy of FeynRules is to allow the model to be written in a general enough way to allow translation into any Feynman diagram generator.  For this reason, when FeynRules loads a model it stores the information in a generic way that can be used to write model files for any Feynman diagram calculator.  This translation is done by interfaces.  A handful of interfaces are already written, but we would like to see many more.  For this reason, we include details about how the model information is stored and how to write an interface to your favorite Feynman diagram calculator.  We also include a template interface as described later in this section.

The information about the model is primarily stored in five lists: {\tt PartList}, {\tt MassList}, {\tt WidthList}, {\tt EParamList} and {\tt IParamList}.  To write an interface, the user creates a set of Mathematica functions that take the information in these lists and write them to files with the appropriate format for their diagram calculator.  In this section, we describe each of these lists and conclude with a description for how to use these lists to make an interface.

\subsection*{The Particle List}
The particle list is called {\tt PartList} and contains a list of all the particles of the model with their properties.  Each element of {\tt PartList} is a list which describes one particle class.  We will describe the elements of this list with an example.  Here is one element of {\tt PartList} for the charged $+2/3$ quarks in the SM:
\begin{verbatim}
{ {F[3],uq} , {
    {"u", "u~", F, S, ZERO, ZERO, T, "u", 2, 
                 "u-quark", "u", "u~", NoGS},
    {"c", "c~", F, S, MC, ZERO, T, "c" ,4 ,
                 "c-quark'', "c", "c~", NoGS},
    {"t", "t~", F, S, MT, WT, T, "t", 6,
                 "t-quark", "t", "\bar{t}", NoGS},
    }
}
\end{verbatim}
The first component of this list contains the class name as a two component list.  The first component is the class type and number (F[3] in this example) while the second component is the class name (uq in this example).  The second component of this class list is a list of the class members.  There are three charged $+2/3$ quarks, so there are three elements of this list.  Each element is a list of the properties of a class member. In the following we give a small description of the contents of this list.
\begin{enumerate}
\item The first element of the class member list refers to the particle names as given by the {\tt ParticleName} option.  In this case, they are \verb+"u"+, \verb+"c"+ and \verb+"t"+.  
\item The second element gives the antiparticle name as given by the {\tt AntiPart- icleName} option.  In this case, they are  \verb+"u~"+, \verb+"c~"+ and \verb+"t~"+.  
\item The third element specifies the spin of the particle and is determined by the symbol used in the class declaration. In this case they are all fermions and so this element is an {\tt F}.  
\item The fourth element refers to the type of line to use for the propagator as given by the {\tt PropagatorType} option. For this quark, this is an {\tt S} for straight line.  
\item The fifth element gives the name of the mass for the particle. The u-quark is massless while the c-quark and t-quark masses are given by the names {\tt MC} and {\tt MT} respectively.  
\item The sixth element is the width symbol of the particle.  
\item The seventh element gives the color representation of the particle. In this case, the particles are labeled {\tt T} for triplet or fundamental.  
\item The eighth element is the {\tt PropagatorLabel}. 
\item The ninth element is the PDG number of the particle, in this case 2, 4, and 6.  
\item The tenth element is a string giving the full name of the particle as given by the {\tt FullName} option.  
\item[(11,12)] The eleventh and twelfth element are the \TeX{} names for the particle and antiparticle respectively.  
\item[(13)] The thirteenth element determines what gauge boson eats this particle if it is a Goldstone boson.  In this case, the quark is not a Goldstone boson, so it is {\tt NoGS}.
\end{enumerate}

\subsection*{The Mass and Width Lists}
The mass and width lists are called {\tt MassList} and {\tt WidthList} respectively.  They contain a list of the masses and widths of the particles.  Each is a two component list.  The first component is the word ``Mass'' for the mass list and ``Width'' for the width list.  The second component is a list of the masses or widths of each particle in the model.  As an example, consider the mass and width of the top quark.  We begin with the mass which is given in this list as:
\begin{verbatim}
{ {6} , MT , 174.3 }
\end{verbatim}
The first element gives  6 as the PDG number of the top quark, MT gives the name of the top quark mass and 174.3 gives the numerical value of the top quark mass.

On the other hand, its width is given by:
\begin{verbatim}
{ {6} , WT , 1.50834 }
\end{verbatim}
The first element gives 6 as the PDG number of the top quark, WT gives the name of the top quark width and 1.50834 gives the numerical value of the top quark width.

\subsection*{The External Parameter List}
The external parameters are stored in a list called {\tt EParamList}.  It contains a list of the external parameters and properties.  We describe it with an example.  Here is the SMINPUTS block from the SM:
\begin{verbatim}
{SMINPUTS, {
    {{1}, {aEWM1, QED, 2, 127.9, False, 
        "Inverse of the electroweak coupling constant"}}, 
    {{2}, {Gf, QED, 2, 0.000011663900000000002, False, 
        "Fermi constant"}}, 
    {{3}, {aS, QCD, 2, 0.118, False, 
        "Strong coupling constant at the Z pole."}}, 
    {{4}, {ZM, 91.188, False, "Z mass"}}
}}
\end{verbatim}
The first element gives the name of the block, SMINPUTS in this case.  The second element is a list of the parameters in the block.  There are four parameters in this block.  For each parameter, the first element is the number of the parameter in the block, 1, 2, 3 and 4 in this example.  The second element is a list of the properties of the parameter.  The elements of this list are as follows:
\begin{enumerate}
\item  The first element of the properties list is the name of the parameter, we have {\tt aEWM1}, {\tt Gf}, {\tt aS} and {\tt ZM}.  
\item[(2,3)] For the parameters where the {\tt InteractionOrder} option was specified, the second and third elements contain this information.  Thus, {\tt aEWM1} and {\tt Gf} are 2nd order in the {\tt QED} coupling, {\tt aS} is 2nd order in the {\tt QCD} coupling and {\tt ZM} does not have this information specified so these two elements are removed.  
\item[(-3)] The third to last element is the numerical value of the parameter.  {\tt aEWM1} has value {\tt 127.9}, {\tt aS} has value {\tt 0.118} and so on.  
\item[(-2)] The second to last element specifies whether the parameter is complex.  These are all real.  
\item[(-1)] The last element is the name of the parameter as given by the {\tt Description} option, for example \verb+"Z mass"+ for the parameter {\tt ZM}. 
\end{enumerate}

\subsection*{The Internal Parameter List}
The internal parameters are stored in a list called {\tt IParamList}.  It contains a list of the internal parameters and properties.  We explain with an example.  Here is one line of {\tt IParamList} from the SM.  It specifies $\alpha_{EW}$:
\begin{verbatim}
{aEW, aEWM1^(-1), QED, 2, False, "Electroweak coupling constant"}
\end{verbatim}
The elements of this list are:
\begin{enumerate}
\item The first element is the name of the parameter, {\tt aEW}.  
\item The second element is the formula giving its values in terms of other parameters.  In this case, it is the inverse of the external parameter {\tt aEWM1}. 
\item[(3,4)] If the {\tt InteractionOrder} option was specified, the  third and fourth elements contain this information.  This parameter is  2nd order in the {\tt QED} coupling.
\item[(-2)] The second to last element specifies whether this element is complex.  It is real.  
\item[(-1)] The last element gives the name of the parameter as specified in the option {\tt FullName}.  
\end{enumerate}

\subsection*{Writing a Translation Interface}
A translation interface cycles through the information in the lists just described and writes the relevant information to files in the appropriate format.  Each Feynman diagram calculator has its own subtleties that make the job slightly more challenging than this, but that is the basic idea.  We suggest the following minimal set of functions for the interface (where Template should be changed to the name of your format):
\begin{itemize}
\item {\tt WriteTemplateOutput[}$\mathcal{L}_1$, $\mathcal{L}_2$, \ldots, \emph{options} {\tt ]} : This should be the main function a user calls to write a model in the Template format.  It should call the other functions to write each component of the model file.  \emph{options} can be whatever the interface writer deems appropriate.
\item {\tt WriteTemplateParticles[} \emph{options} {\tt ]} : This function should write the particles to the file in the Template format and be used inside {\tt WriteTemplate- Output}.
\item {\tt WriteTemplateExtParams[} \emph{options} {\tt ]} : This should write the external parameters to the file in the Template format.  It should be used both inside {\tt WriteTemplateOutput} and by the user to implement a new set of numerical values without rewriting the entire model.
\item {\tt ReadTemplateExtParams[} \emph{options} {\tt ]} : This should read the external parameters from the file in the Template format and update the parameters in FeynRules accordingly.  This function will not be used inside {\tt WriteTemp- lateOutput}.
\item {\tt WriteTemplateIntParams[} \emph{options} {\tt ]} : This function should write the internal parameters to file in the Template format.  It should be called inside {\tt WriteTemplateOutput}.
\item {\tt WriteTemplateVertices[}$\mathcal{L}_1$, $\mathcal{L}_2$, \ldots, \emph{options} {\tt ]} : This function should be called from {\tt WriteTemplateOutput} and should write the vertices to file in the Template format.
\end{itemize}

In addition to the lists outlined earlier in this section, an interface writer will usually need the list {\tt ParamRules} which contains a replacement list that takes the parameter symbols to the parameter names, as well as the function {\tt PartName}, which takes as an input the \emph{Mathematica} name of a particle and return the corresponding particle name.  This is usually necessary since the parameter and particle symbols can be any Mathematica symbol and may not be compatible with the Feynman diagram program.  The parameter and particle names are intended to be plain characters and should be appropriate for any program.

We have created a template interface in FeynRules.  The interface writer can look at it for guidance and start from scratch, or they can copy it directly to a new name and modify it to fit their needs.  It can be found in the \verb+Subroutines+ directory of FeynRules and is called ``TemplateInterface.m''.

\section{Code Details}
\label{sec:codedetails}

\subsection*{Model File Format}
\label{sec:Model File Format}
As we have mentioned, the FeynRules model file format is an extension of the FeynArts format.  The particles are grouped in classes as in FeynArts in a list variable called {\tt M\$ClassesDeclaration}.  The benefits of using classes in FeynRules are that more compact expressions for the particle definitions, the Lagrangian and the vertices can be accomplished.  The user who implements a model can group together the particles with the same quantum numbers into ``classes'' and then write down the Lagrangian for the entire class of particles at once rather than for each member individually.  This can save time and reduce errors.  Furthermore, the vertices can be generated in a compact form.  One vertex can represent the vertices involving all the members of a class by including flavor indices which span the class members.  It also allows a straight forward translation to FeynArts.  These particle classes have been extended with several options not available in FeynArts.  These options allow the specification of particle properties which are not used in FeynArts but are necessary for translation to other Feynman diagram calculators.  An example is the PDG number of a particle.

We also introduce several new variables to the FeynRules model file while removing one.  Since the model is defined in terms of a Lagrangian in FeynRules, there is no need for the vertices list {\tt M\$CouplingMatrices}.  The vertices are derived from the Lagrangian and stored in an internal list for later use.  On the other hand, the FeynArts model file did not have variables for storing definitions of parameters or gauge groups.  For this reason, the lists {\tt M\$Parameters} and {\tt M\$GaugeGroups} were defined.  In analogy to the parameters, it turned out convenient to allow the user to define parameter and gauge group ``classes'' with properties.  For parameters, this includes the ability to collect parameters that act on different members of a particle class into one class of parameters with a flavor index.  An example of this is the CKM matrix.  

By using the concept of classes for the definitions of the particles, gauge groups and parameters in the model file, a unified approach to initialization of the model was allowed.  When a model is loaded, these definition lists are read in.  After reading the model file into memory, FeynRules goes through each list one class definition at a time.  For each class definition, FeynRules first saturates the options with their default values (where none were specified by the user).  It then stores the information in a set of global variables that can be used later (\eg by interface writers).  These lists are as follows:
\begin{itemize}
\item[-] {\tt PartList} : This is a list of the physical\footnote{By physical we mean the classes for which the option {\tt Unphysical} was not set to {\tt True}.  Typically, this means that they are the mass eigenstates.} particle classes, their members and their properties.
\item[-] {\tt MassList} : This is a list of the masses of the particles and their properties.
\item[-] {\tt WidthList} : This is a list of the widths of the particles and their properties.
\item[-] {\tt EParamList} : This is a list of the external parameters and their properties.
\item[-] {\tt IParamList} : This is a list of the internal parameters and their properties.
\end{itemize}
Further details of the contents of these lists can be found in Section \ref{sec:Implementing an Interface}.  Together with the Feynman rules, this information is sufficient to write a model file for a Feynman diagram calculator.

\subsection*{Index Restoration}
In order to derive the Feynman rules, all the indices need to be restored and appropriately contracted.  We do this by first restoring all the indices, according to the index ordering defined in the model file.  FeynRules next contracts the indices of the same type between nearest neighbor indices.  It does this first on the nonfield terms (parameters and gamma matrices).  

For example, consider the quark, W boson vertex from the SM.  It contains the CKM matrix defined on page \pageref{CKM definition}.  Its Lagrangian term looks like:
\begin{verbatim}
uqbar.Ga[mu].ProjM.CKM.dq W[mu]
\end{verbatim}
where we have removed the overall constants.  We begin by restoring the indices for the gamma matrices and CKM matrix according to the definitions, giving:
\begin{verbatim}
uqbar.Ga[Index[Lorentz,mu],Index[Spin,s1],Index[Spin,s2]].
     ProjM[Index[Spin,s3],Index[Spin,s4]].
     CKM[Index[Generation,a1],Index[Generation,a2]].dq 
     W[Index[Lorentz,mu]]
\end{verbatim}
Notice that if a tensor represents a matrix, then the last two indices refer to the matrix indices, \eg $\gamma^\mu_{sr}$ is represented by {\tt Ga[mu,s,r]}, and not {\tt Ga[s,mu,r]}. FeynRules then cycles through each index type and contracts nearest neighbors. The left and right most indices are contracted to the fields at the corresponding end of the chain. Since the indices are of different type, there is no ambiguity. This gives:
\begin{verbatim}
TensDot[Ga[mu], ProjM][Index[Spin,s1],Index[Spin,s4]] 
       CKM[Index[Generation,a1],Index[Generation,a2]] 
       uqbar[Index[Spin, s1], Index[Generation, a1]].
            dq[Index[Spin, s4], Index[Generation, a2]] 
       W[Index[Lorentz,mu]]
\end{verbatim}
Two comments are in order about this last expression:
\begin{enumerate}
\item As all the tensors are now contracted to the fields at the end of the chain, the resulting matrix elements can now be treated as numbers and moved outside the chain built by the {\tt Dot} product.
\item In order to avoid a proliferation of interconnected indices in matrix products, the resulting matrix products are wrapped inside the {\tt TensDot} function.  For example, instead of writing {\tt Ga[mu,s1,s2]ProjM[s2,s4]}, FeynRules replaced this with {\tt TensDot[Ga[mu],ProjM][s1,s4]}.
\end{enumerate}
This takes care of all the indices which are contracted with one or more tensors.  Finally, the indices left over on the fields which do not connect to tensors are contracted with each other in pairs.  

\subsection*{Feynman Rules}
The Feynman rules are generated using the rules of canonical quantization.  The way this is done in FeynRules is the following.  Suppose we are given the Lagrangian term:
\begin{equation}
\mathcal{L} = g_{\alpha_1\cdots\alpha_n}\partial\cdots\partial\ \phi_{\alpha_1}\cdots\phi_{\alpha_n}
\end{equation}
where $\alpha_i$ represents all the indices and quantum numbers of the field $\phi_{\alpha_i}$ and $\partial$ is a derivative acting on one or more of the fields.  This Lagrangian term is multiplied on the right by the creation operators for the fields (where we take all momenta to be ingoing).  The order of the fermionic and ghost creation operators is the reverse of the order of the fermions and ghosts in the Lagrangian term.  Once this is done, the product is sandwiched between the vacuum giving:
\begin{equation}
g_{\alpha_1\cdots\alpha_n}
\partial\cdots\partial\ 
\left<0\left|
\phi_{\alpha_1}(x)\cdots\phi_{\alpha_n}(x)
a_{\bar{\phi}_{\alpha_n}}^\dagger(p_n)\cdots a_{\bar{\phi}_{\alpha_1}}^\dagger(p_1)
\right|0\right>
\end{equation}
The creation operators are then moved to the left using the (anti)commutation rules of the fields.
\begin{equation}
\left[\phi_{\alpha_i}(x),a_{\bar{\phi}_{\alpha_j}}^\dagger(p)\right] = 
\delta_{\alpha_i\alpha_j}u_{\alpha_i}(p)e^{-ipx}
\end{equation}
where $u_{\alpha_i}$ is the wavefunction of $\phi_{\alpha_i}$.  When they get to the left end, they annihilate the vacuum.

Once all the creation operators are gone, the vacuum states act on each other giving one.  Any derivatives are now applied and pull down factors of momentum from the exponentials.  Finally, the $Exp[-i(p_1+\cdots+p_n)x]$ and the wavefunctions $u_{\alpha_i}$'s are dropped and the result is multiplied by $i$ giving the Feynman rule vertex.

We  illustrate this procedure on the QED interaction given by:
\begin{equation}
\mathcal{L}_{\bar{\psi}\psi A}= - e  \bar{\psi}\gamma^\mu \psi A_\mu
 = -e \bar{\psi}_{s,f}\gamma^{\mu}_{ss'} \psi_{s',f} A_\mu,
\end{equation}
where $e$ denotes the electromagnetic coupling constant, $\psi_f$ the lepton field which may have one of three flavors ($e,\mu,\tau$) and $A$ the photon field.

This term is multiplied on the right by $a^\dagger_{\bar{\psi}}a^\dagger_{\psi}a^\dagger_{A}$ and sandwiched between the vacuum to give:
\begin{equation}
\label{eq:matrixelement}
-e\langle0|\bar{\psi}\gamma^\mu \psi A_\mu a^\dagger_{\bar{\psi}}a^\dagger_{\psi}a^\dagger_{A}|0\rangle,
\end{equation}
where $a^\dagger_{\bar{\psi}}$, $a^\dagger_{\psi}$ and $a^\dagger_{A}$ denote the creation operators associated to the fields $\psi$, $\bar{\psi}$ and $A$ in the canonical formalism.

We next move the creation operators all the way to the left using the (anti) commutation relations:
\begin{eqnarray}
\left\{\ \psi_{s,f}(x)\ ,\ a^\dagger_{\bar{\psi}_{s',f'}}(p)\ \right\} &=&
  \delta_{ss'}\delta_{ff'}u_{s,f}e^{-ipx}\\
\left\{\ \bar{\psi}_{s,f}(x)\ ,\ a^\dagger_{\psi_{s',f'}}(p)\ \right\} &=&
  \delta_{ss'}\delta_{ff'}\bar{u}_{s,f}e^{-ipx}\\
\left[\ A_\mu(x)\ ,\ a^\dagger_{A_\nu}(p)\ \right] &=&
  \delta_\mu^\nu \epsilon_\mu e^{-ipx}
\end{eqnarray}
which results in:
\begin{equation}
-e\delta_{ff'}\gamma^\mu_{ss'}\bar{u}_{sf}(p_1) u_{s'f'}(p_2) \epsilon_\mu(p_3) e^{-i(p_1+p_2+p_3)x}
\end{equation}

Finally, dropping the wave functions, the $Exp[-i(p_1+p_2+p_3)x]$ and multiplying by $i$ gives the well known QED vertex
\begin{equation}
-ie\delta_{ff'}\gamma^\mu_{ss'}.
\end{equation}

\subsection*{Majorana Fermions and Conjugated Fermions}
Particular care is needed when dealing with interaction terms involving Majorana fermions or explicit charge conjugation. Feynman rules involving Majorana fermions and explicit charge conjugation are handled using the fermion flow algorithm of Ref.~\cite{Denner:1992vza}.We show here how Majorana fermions are treated in the code, charge conjugation is dealt with in a similar way. 

Let us consider the following fermion chain
\begin{equation}
\label{eq:MajLag}
\bar{\lambda}_1\Gamma\lambda_2,
\end{equation}
where $\lambda_1$ and $\lambda_2$ denote Majorana fermions, and $\Gamma$ denotes a chain of Dirac matrices.  We have to distinguish two situations
\begin{enumerate}
\item If $\lambda_1= \lambda_2=\lambda$, we need to take into account the fact that a Majorana fermion is a self conjugate field, \ie
\begin{equation}
\{\bar{\lambda}(x), a^\dagger_{\bar{\lambda}}(p)\} = e^{i\delta}\bar{v}(p) e^{-ipx},
\end{equation}
where $\delta$ is the phase carried by the Majorana fermion.
Furthermore, we have to ``symmetrize'' Eq.~(\ref{eq:MajLag}) to take into account relations like $\bar{\lambda}\gamma^\mu\lambda=0$ for Majorana fermions.
The symmetrization is done by calculating the reversed fermion flow, defined by
\begin{equation}
\bar{\psi}_1\Gamma\psi_2 = \kappa_i\bar{\psi_2^c}\tilde\Gamma\psi_1^c,
\end{equation}
where $\kappa_i=\pm1$ and $\tilde\Gamma$ denotes the reversed chain of Dirac matrices. (For example, if $\Gamma=\gamma^{\mu_1}\gamma^{\mu_2}\cdots\gamma^{\mu_n}$ then $\tilde{\Gamma}=\gamma^{\mu_n}\cdots\gamma^{\mu_2}\gamma^{\mu_1}$.)  The symmetrization is then performed via
\begin{equation}
\bar{\lambda}\Gamma\lambda=\frac{1}{2}\bar{\lambda}\Gamma\lambda+\frac{1}{2} \kappa_i\bar{\lambda}\tilde\Gamma\lambda,
\end{equation}
where we used the fact that $\lambda^c=e^{i\delta}\lambda$. Applying the symmetrization formula implies for example
\begin{equation}
\bar{\lambda}\gamma^{\mu}P_-\lambda =  \frac{1}{2}\bar{\lambda}\gamma^{\mu}P_-\lambda - \frac{1}{2}\bar{\lambda}\gamma^{\mu}P_+\lambda =-\bar{\lambda}\gamma^{\mu}\gamma^5\lambda.
 \end{equation}
 \item
 If $\lambda_1\neq\lambda_2$, we do not need to symmetrize the expression, but we need to be careful to use the same fermion flow for all contributions to the vertex:
 \begin{equation}
a \bar{\lambda}_1\Gamma_i\lambda_2 + b \bar{\lambda}_2\Gamma_j\lambda_1=a \bar{\lambda}_1\Gamma_i\lambda_2 + \kappa_jb\, e^{i(\delta_2-\delta_1)}\bar{\lambda}_1\tilde\Gamma_j\lambda_2.
\end{equation}
In this last expression we have the same ordering for the chain ($\lambda_1, \lambda_2$), and so we can consistently extract the corresponding vertex.
 \end{enumerate}

Similar transformations are applied if the fermion chain contains explicit charge conjugated fields. These transformations allow us to define the fermion flow consistently for each fermion chain. The convention followed in FeynRules for the fermion flow of the vertex ($X$, $\bar{\psi_1}$, $\psi_2$), where $X$ denotes either a scalar or a vector field, is that $\psi_2$ is considered ingoing and $\psi_1$ outgoing. As an illustration we show in Fig.~\ref{fig:cflow} the two fermion flows contributing to a fermion number violating vertex, and how the two flows are represented in the FeynRules output\footnote{All Feynman diagrams were drawn using Jaxodraw~\cite{Binosi:2003yf}.}.
\begin{center}
\begin{figure}[!t]
\fcolorbox{white}{white}{
  \begin{picture}(344,66) (15,-30)
    \SetWidth{0.5}
    \SetColor{Black}
    \ArrowLine(49,1)(107,1)
    \ArrowLine(153,24)(107,1)
    \Photon(107,1)(153,-22){3.5}{5}
    \ArrowArc(79.01,94.95)(87.41,-95.91,-51.02)
    \Text(36,1)[lb]{\Large{\Black{$\psi_2$}}}
    \Text(155,-30)[lb]{\Large{\Black{$X$}}}
    \Text(156,20)[lb]{\Large{\Black{$\psi_1^c$}}}
    \Text(15,1)[lb]{\Large{\Black{$a)$}}}
    \ArrowLine(222,1)(280,1)
    \ArrowLine(326,24)(280,1)
    \Photon(280,1)(326,-22){3.5}{5}
    \ArrowArc(251.72,95.44)(87.98,-96.34,-51.07)
    \Text(209,1)[lb]{\Large{\Black{$\psi_1$}}}
    \Text(327,-30)[lb]{\Large{\Black{$X$}}}
    \Text(329,20)[lb]{\Large{\Black{$\psi_2^c$}}}
    \Text(188,1)[lb]{\Large{\Black{$b)$}}}
  \end{picture}
}
\caption{\label{fig:cflow}The two fermion flows contributing to a vertex containing an explicit charge conjugation: a) the vertex ($X$, $\bar{\psi_1^c}$, $\psi_2$), b) the vertex ($X$, $\bar{\psi_2^c}$, $\psi_1$).}
\end{figure}
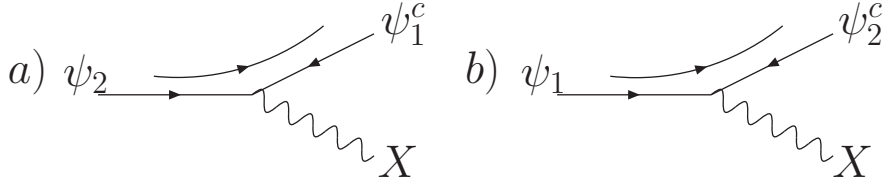
\end{center}

The same rule holds true for interactions involving Majorana fermions. The two fermion flows of the vertex ($X$, $\lambda_1$, $\lambda_2$) are shown in Fig.~\ref{fig:mflow}.  Notice that in this context the ``bar'' on the Majorana fermion serves only to define the direction of the fermion flow, and does not refer to an antiparticle, \ie the vertex corresponding to the fermion flow where $\lambda_2$ is ingoing and $\lambda_1$ is outgoing is written in the FeynRules output as ($X$, $\bar{\lambda}_1$, $\lambda_2$), whereas the reversed flow where $\lambda_1$ is ingoing and $\lambda_2$ is outgoing is written as ($X$, $\bar{\lambda}_2$, $\lambda_1$).
\begin{center}
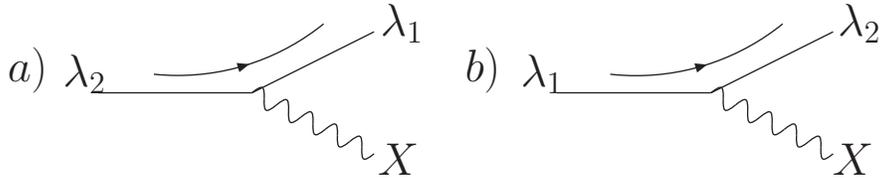
\begin{figure}[!t]
\fcolorbox{white}{white}{
  \begin{picture}(344,66) (15,-30)
    \SetWidth{0.5}
    \SetColor{Black}
    \Line(222,1)(280,1)
    \Line(326,24)(280,1)
    \Photon(280,1)(326,-22){3.5}{5}
    \ArrowArc(251.72,95.44)(87.98,-96.34,-51.07)
    \Text(209,1)[lb]{\Large{\Black{$\lambda_1$}}}
    \Text(327,-30)[lb]{\Large{\Black{$X$}}}
    \Text(329,20)[lb]{\Large{\Black{$\lambda_2$}}}
    \Text(188,1)[lb]{\Large{\Black{$b)$}}}
    \Line(49,1)(107,1)
    \Line(153,24)(107,1)
    \Photon(107,1)(153,-22){3.5}{5}
    \ArrowArc(79.01,94.95)(87.41,-95.91,-51.02)
    \Text(36,1)[lb]{\Large{\Black{$\lambda_2$}}}
    \Text(155,-30)[lb]{\Large{\Black{$X$}}}
    \Text(156,20)[lb]{\Large{\Black{$\lambda_1$}}}
    \Text(15,1)[lb]{\Large{\Black{$a)$}}}
  \end{picture}
}
\caption{\label{fig:mflow}The two fermion flows contributing to a vertex containing two Majorana fermions: a) the vertex ($X$, $\bar{\lambda_1}$, $\lambda_2$), b) the vertex ($X$, $\bar{\lambda_2}$, $\lambda_1$).}
\end{figure}
\end{center}

\subsection*{Implemented models and validation}
In order to validate our code, we implemented several models into FeynRules, and compared the Feynman rules obtained by our code with the interaction vertices given in the literature. We furthermore implemented these models into Monte Carlo programs using the corresponding FeynRules interfaces, and checked that for each of them the cross-section we obtain agrees with the stock version of the Monte Carlo program. A more detailed paper discussing the different Monte Carlo interfaces and their validation will be published in the near future, so we will be brief on this in the current paper.

All the models described in this section can be downloaded from the FeynRules web site. These models represent ``base models'', which can serve as a starting point when writing a new model. Indeed, in many cases it is not necessary to write a new model file from scratch where the new model is an extension of an already existing model implementation (\eg MSSM and NMSSM). In the rest of this section we give a short review of the models already implemented, as well as the checks we did for their validation.

We implemented the complete Standard Model (SM), and checked the Feynman rules that we obtained to those given in the literature, both in unitary and in Feynman gauge. We found complete agreement for all vertices. We furthermore implemented the SM in FeynArts using the corresponding FeynRules interface and checked that all the couplings given in \verb+M$CouplingMatrices+\comment{$} agree with the ones given in the default FeynArts model file for the SM. Finally, we checked that our SM implementation agrees with the cross-sections obtained by the default MadGraph/MadEvent and CalcHEP/CompHEP implementation on a selection of 22 processes, both in unitary and in Feynman gauge\footnote{In MadGraph/MadEvent only unitary gauge is supported.}.

We also implemented the Three-Site Model\cite{SekharChivukula:2006cg} a minimal Higgsless model.  We checked our implementation against the existing CalcHEP implementation\cite{He:2007ge} on 189 key processes in both Feynman and unitary gauge and on both CalcHEP and CompHEP.

Finally, implementations of the the MSSM and a model of minimal extra dimension where performed by B.~Fuks and P.~Aquino. The authors of the model files checked that the vertices obtained by FeynRules agree with the known textbook expressions. The validation of these models using the Monte Carlo and FeynArts/FormCalc interfaces is currently ongoing.
\section{Conclusions}
\label{sec:conclusion}
In this paper we presented FeynRules, a \emph{Mathematica} package to extract Feynman rules from a Lagrangian, and which allows to implement new physics models into several Feynman diagram calculators in an automated way. We explained how to write a model file, the key ingredient to implement a BSM model
into FeynRules. The model file, together with the Lagrangian describing the model, can then be used to derive the interaction vertices. The model information as well as the vertices computed by FeynRules are stored inside \emph{Mathematica} in a generic way, which gives the possibility to write out model files in any format of choice. This enables us to write translation interfaces which transform the generic model information into the specific format of a given Feynman diagram calculator. Such interfaces are currently available for CalcHEP/CompHEP, FeynArts/FormCalc, MadGraph/MadEvent and Sherpa, but we hope that we can extend this list in the future, in order to cope with the challenge given by the LHC.
\section*{Acknowledgments}
The authors would like to thank S.~Chivukula and T.~Hahn for useful discussions, as well as R.~Frederix, M.~Herquet, F.~Maltoni and S.~Schumann for help with the MC interfaces. We would also like to thank P.~Aquino, M.~Feliciangeli, B.~Fuks for their help with the debugging of the code and the implementation of new models. CD is a research fellow of the ``Fonds National de la Recherche Scientifique'', Belgium.  NDC was supported by the US National Science Foundation under grants PHY-0354226 and PHY-0555544.

\bibliography{FRBib}

\end{document}